\documentclass[a4paper,11pt]{article}
\pdfoutput=1 

\usepackage{jheppub} 
\usepackage[utf8]{inputenc}
\usepackage{setspace}

\usepackage{hyperref}
\definecolor{refkey}{gray}{0.45}
\definecolor{labelkey}{RGB}{155,48,48}
\definecolor{UI_blue}{RGB}{32, 64, 151}
\definecolor{UI_red}{RGB}{187, 62, 24}
\definecolor{UI_blue2}{RGB}{0, 84, 147}
\definecolor{UI_red2}{RGB}{159, 32, 66}
\definecolor{UI_gray}{RGB}{169, 169, 169}
\definecolor{UI_sepia}{RGB}{112, 66, 20}
\definecolor{UI_bittersweet}{RGB}{254, 111, 94}
\definecolor{UI_emerald}{RGB}{80, 200, 120}
\definecolor{UI_olivegreen}{RGB}{181, 179, 92}
\definecolor{UI_cadetblue}{RGB}{95, 158, 160}
\definecolor{UI_fuchsia}{RGB}{255, 0, 255}
\definecolor{UI_midnightblue}{RGB}{25, 25, 112}
\definecolor{UI_royalblue}{RGB}{0,35, 102}
\definecolor{UI_periwinkle}{RGB}{204, 204, 255}
\definecolor{UI_redorange}{RGB}{255, 83, 73}
\definecolor{UI_brickred}{RGB}{203,65,84}	
\definecolor{UI_forestgreen}{RGB}{34, 139, 34}
\definecolor{UI_tan}{RGB}{210,180,140}	
\definecolor{UI_burlywood}{RGB}{222,184,135}
\definecolor{UI_burlywood}{RGB}{192,64,0}
\definecolor{UI_darkorchid}{RGB}{153,50,204}

\usepackage{mathtools}
\usepackage{physics}
\usepackage{simplewick}
\usepackage[obeyFinal]{todonotes}

\def\beq{\begin{eqnarray}}\def\eeq{\end{eqnarray}}
\def\be{\begin{equation}}\def\ee{\end{equation}}

\def\mes[#1]{d^{3}{#1}}

\reversemarginpar

\title{A CFT interpretation of cosmological correlation functions in $\alpha-$vacua in de-Sitter space}

\preprint{\parbox{3cm}{TIFR/TH/22-32}}

\author[a]{Sachin Jain,}
\author[b]{Nilay Kundu,}
\author[c]{Suman Kundu,}
\author[a]{Abhishek Mehta,}
\author[c]{Sunil Kumar Sake}

\affiliation[a]{Indian Institute of Science Education and Research, Homi Bhabha Road, Pashan, Pune 411 008, India}
\affiliation[b]{Department of Physics, Indian Institute of Technology Kanpur, Kalyanpur, Kanpur 208016, India}
\affiliation[c]{Department of Theoretical Physics, Tata Institute of Fundamental Research, Colaba, Mumbai, India, 400005}

\emailAdd{sachin.jain@iiserpune.ac.in}
\emailAdd{nilayhep@iitk.ac.in}
\emailAdd{suman.kundu\_290@tifr.res.in}
\emailAdd{abhishek.mehta@students.iiserpune.ac.in}
\emailAdd{sunil.sake@tifr.res.in}

\vspace{1cm}

\abstract{ de-Sitter(dS) space allows for a generalized class of vacua, known as $\alpha-$vacua, described by some parameters.  The Bunch-Davies (BD) vacuum is a point in this parameter space. The 
	cosmological correlation function  in  BD vacuum in four dimensions and can be interpreted as  $CFT_3$ correlation function of certain operators. However, the correlation function in $\alpha-$vacua takes a much more complicated form. In this paper, we  give a simple prescription to compute correlation function in $\alpha-$vacua in terms of correlation function of BD vacuum. We also show that the correlation function in the $\alpha-$vacua can be related to three-dimensional CFT correlation functions if we relax the requirement of consistency with  OPE limit. Relaxation of consistency with OPE limit can be naturally achieved in momentum space.}

\begin{document}
	\maketitle
	\flushbottom
	\vskip 10pt
	
	\vskip 20pt
	\section{Introduction}Understanding initial conditions in cosmology is directly related to understanding the primordial universe. Any small deviation in gaussianity in initial conditions is of great interest and will lead to cosmological non-gaussianity. Because of this, theoretical characterization and measurement of non-gaussianity is a very important topic of research.
	Cosmological three point correlation functions measures non-gaussianities \cite{Maldacena:2002vr,Maldacena:2011nz,Komatsu:2010hc}. These correlation functions are mostly computed in Bunch-Davies (BD) vacuum. It turns out that these correlation functions in BD vacuum is same as correlation function of appropriate operators in three dimensional  CFT in momentum space. This is very powerful relation because it provides an  efficient way to calculate the cosmological correlation functions \cite{McFadden:2010vh,Antoniadis:2011ib,McFadden:2011kk,Creminelli:2011mw,Bzowski:2011ab,Mata:2012bx,Ghosh:2014kba,Kundu:2014gxa,Kundu:2015xta}, for recent exciting development see \cite{Arkani-Hamed:2015bza,Arkani-Hamed:2018kmz,Baumann:2019oyu,Baumann:2020dch,Baumann:2021fxj,Baumann:2022jpr,Goodhew:2020hob,Jazayeri:2021fvk,Bonifacio:2021azc,Cabass:2021fnw}. 
	dS space in general allows for a more general class of vacua \cite{Mottola:1984ar, Allen:1985ux} which are related to BD vacuum by Bogoliubov transformation. Their phenomenological significance has been discussed and they are very relevant from observational perspective \cite{Danielsson:2002qh,Goldstein:2002fc,Kanno:2022mkx}. However, there has been a lot of debate if $\alpha-$vacua are at all good for the inflationary scenario as there are several peculiar properties associated with these vacua reported in the literature. For free theory in dS, the Green's function has some nonlocal singularity, more precisely Green's function has singularity between antipodal points which is non-local in nature \cite{Bousso:2001mw}. Moreover, for interacting theory, it is argued that one needs a counter-term that is non-local in nature \cite{Banks:2002nv, Einhorn:2002nu}. It was also shown that the expectation value of stress tensor of probe scalar field  diverges in $\alpha-$vacua in rigid dS space which renders the probe approximation invalid. However, in \cite{Goldstein:2003ut}, it was shown that to renormalize the correlation function, the counter-term requirement is the same as that of BD vacuum.  Interestingly it was also shown that $\alpha-$vacua results are perfectly consistent with causality. For stress tensor, the story is similar and one requires counter-terms which are dependent on parameters that define $\alpha-$vacua. In this paper, we have nothing to add to this story. We continue to assume that $\alpha-$vacua makes sense and compute the cosmological correlation function in this vacua.
	
	Cosmological correlation function for $\alpha-$vacua has been previously computed for scalar fields \cite{Xue:2008mk,Shukla:2016bnu}. These results are much more complicated than that of the Bunch-Davies vacuum. For example, if we consider the three-point function of scalar operators, BD vacuum has only one structure whereas for  $\alpha-$vacua we have more structures and are described by two parameters. One natural question that arises is, what is the CFT interpretation of $\alpha-$vacua correlation functions. In \cite{Bousso:2001mw}, it was argued from dS/CFT perspective that $\alpha-$vacua can be thought of as just a one-parameter family of deformation of usual vacuum by marginal operator deformation, see also \cite{Nguyen:2017ggc} for discussion on construction of analogue of $\alpha-$vacua in CFT  from the universal sector of AdS/CFT dualities. However, in this paper, we are not thinking from dS/CFT perspective. Rather we aim to understand the nature of the $\alpha-$vacua correlation function from the CFT correlation function perspective. 
	
	In this paper, we show that the momentum space CFT correlation function plays a very important role in understanding the $\alpha-$vacua result. In momentum space, the solution to conformal ward identity gives a much richer class of solutions than the position space analog. For example, the solution to momentum space conformal ward identity for the scalar three-point function is given by four solutions\footnote{In position space we have only one structure for scalar three-point function. } \cite{Bzowski:2013sza}.
	In general, this implies that in momentum space we can have a four-parameter space of CFT correlation functions. Further, the requirement of permutation symmetry fixes the momentum space CFT correlator starting from four parameters to two parameters. 
	Interestingly it turns out that these two-parameter space of correlation functions are precisely related to the two-parameter space of correlation function obtained by computing cosmological correlation function in dS space in $\alpha-$vacua. One important distinguishing feature of these $\alpha-$vacua correlation functions as compared to BD vacuum is that these correlators not only have total energy poles for three-point functions but also have other poles which are generally termed as bad poles \footnote{For three-point function in momentum space, it can be shown that only pole structures that are consistent with position space correlation function are total energy poles \cite{Maldacena:2011nz, Shukla:2016bnu, Baumann:2019oyu}. Other pole structures which are generally termed bad poles would lead to singularities that won't be consistent with OPE limit. Interestingly the BD vacua correlation function can be obtained from the general momentum space correlation function by allowing just the total energy pole structure which is equivalent to demanding consistency with OPE limit \cite{Bzowski:2013sza, Maldacena:2011nz, Jain:2021whr}. See \cite{Gillioz:2021sce} for a causality based argument.}. The case of the spinning correlators is even more interesting. It turns out that for the three-point function of conserved charges, even though position space allows for only two parity even correlation functions, the general solution to conformal ward identity in momentum space has many more solutions \cite{Maldacena:2011nz, Jain:2021vrv}. Spinor helicity variables allow for an even larger class of solutions. In this paper, we show that these extra solutions play important role in describing the $\alpha-$vacua correlation function. Interestingly,  it was shown in \cite{Jain:2022ujj} that, these extra solutions in spinor helicity variables also plays important role in relation to flat space amplitude.
	

	Our strategy in this paper can be summarised as follows. 
	\begin{itemize}
		
		\item We first calculate the three-point function of dimension  $\Delta=3$ scalar operator $O_3$ by solving conformal ward identity and constrain the solution by demanding permutation symmetry. This gives us a correlation function that depends on two parameters. By comparing with computation done in a two-parameter family of $\alpha-$vacua in rigid dS space, we identify the two unknown coefficients in CFT in terms of these two parameters. 
		
		\item To compute the spinning correlation function, we simply act on the scalar correlator with spin and weight raising operators.  We cross-check our result by direct computation of spinning correlator in $\alpha-$vacua in rigid dS using in-in formalism.
		
		
		\item We also compute the spinning correlation function by solving conformal ward identity in momentum space and also in spinor helicity variables. It turns out that we have a much larger set of solutions in momentum as well as spinor helicity variables. Demanding consistency with OPE gives the correct BD vacuum answer. However, if we relax consistency with OPE limit and just impose permutation symmetry, for the spinning correlator we still have a larger class of solutions. We show that all these solutions play important role in understanding the correlation function in generalized $\alpha-$vacua\footnote{Notion of generalized alpha vacuum will be defined in section \ref{mxdgnl1}.}.
		
		\item We also give a simple prescription to compute $\alpha-$vacua correlation functions interms of BD vacuum.
	\end{itemize}
	
	The rest of the paper is organized as follows.
	In section \ref{gravta} we review basic facts about $\alpha-$vacua and then calculate the scalar and spinning three-point function using in-in formalism. In section \ref{spinCFT} we demonstrate how the results obtained in $\alpha-$vacua can be obtained from CFT perspective.  In section \ref{generalsol} we show that conformal ward identity allows for a more general class of solutions. We also discuss their relation to correlation functions in generalized $\alpha-$vacua. In section \ref{podd} we discuss how to obtain parity odd correlation function contribution for $\alpha-$vacua using CFT results only. Next, in section \ref{albdrl} we show that correlation functions in $\alpha-$vacua can be expressed in terms of the same answer in BD vacuum. In other words, we will show that given the structure in BD vacuum, we can obtain the corresponding expression for the correlator in $\alpha-$vacuum. In section \ref{dis} we summarise the findings of this paper.
	In Appendices, we collect some useful background details which are helpful in the main draft.

	\section{Cosmological correlation function in $\alpha-$vacua}\label{gravta}
	The aim of this section is to use the in-in formalism to calculate cosmological correlation functions. We first start with a brief review of $\alpha-$vacua. We then use in-in formalism to calculate scalar and spinning three-point functions.
	
	We work in poincare coordinate in dS. The metric is given by
	\begin{equation}
		ds^2=\frac{1}{H^2\eta^2}\left(-d\eta^2 + dx^i dx_{i}\right)
	\end{equation}
	with $-\infty<\eta\le 0$ where $H$ is Hubble parameter, which we set to unity in this paper. Let us consider free  scalar field theory 
	\begin{equation}\label{msds1}
		S=-\frac{1}{2} \int d^3x d\eta \sqrt{-g} g^{\mu\nu} \partial_{\mu}\varphi \partial_{\nu}\varphi.
	\end{equation}
	Mode expansion of the scalar field in Bunch-Davies vacuum is given by
	\begin{equation}\label{modesc}
		\varphi(x,\eta)=\int \frac{d^3k}{\left(2\pi\right)^3} \left(a_{k} v_{k}(\eta)+ a_{-k}^{\dagger} v^{*}_{k}(\eta) \right)e^{ik.x}
	\end{equation}
	where $v_{k}(\eta)=\frac{1}{\sqrt{2 k^3}} \left(1+ i k \eta \right) e^{- i k \eta}.$
	The Bunch-Davies vacuum is defined by
	\begin{equation}
		a_{k} | 0\rangle =0\,\,\forall \,\, k  
	\end{equation}
	For dS space, one can define two real parameter set of general vacuum $| \alpha, \beta\rangle$\cite{Allen:1985ux}.
	The mode expansion of massless scalar field \eqref{msds1} $\varphi$ is given by
	\begin{align}
		\varphi(\eta, \boldsymbol{x})=\int \frac{d^{3} k}{(2 \pi)^{3}}\left[b_{\boldsymbol{k}} u_{k}(\eta)+b_{-\boldsymbol{k}}^{\dagger} u_{k}^{*}(\eta)\right] \mathrm{e}^{i \boldsymbol{k} \cdot \boldsymbol{x}}\label{modex}
	\end{align}
	and
	\begin{align}
		u_{k}(\eta)=\frac{1}{\sqrt{2 k^{3}}}\left\{A(1-i k \eta) \mathrm{e}^{i k \eta}+B(1+i k \eta) \mathrm{e}^{-i k \eta}\right\}\label{me}
	\end{align}
	where $A, B$ are arbitrary complex numbers which satisfies
	\begin{equation}\label{conAB}
		|A|^2-|B|^2=1.
	\end{equation}
	One choice of parameters is 
	\begin{equation}\label{absca}
	A= \cosh\left(\alpha\right),~~~~	B= - i e^{i\beta} \sinh\left(\alpha\right).
	\end{equation}
	The $| \alpha, \beta\rangle$ vacuum is defined by
	\begin{equation}\label{albtvc}
		b_{k} | \alpha, \beta\rangle =0\quad \forall \,\, k
	\end{equation}
	Let us not that BD vacuum is a special case $\alpha=0$ of this general vacuum. One can write down the following relation
	\begin{equation}
		| \alpha, \beta\rangle = \prod_{k} \frac{1}{\sqrt{|B|}}  \exp\left( \frac{A^{*}}{2 B^*} a_{k}^{\dagger}a_{-k}^{\dagger}\right).
	\end{equation}

	\subsection*{Mode expansion for metric fluctuations}
	We take the mode expansion in $\alpha-$vacua for the metric fluctuation to be given by
	\begin{equation} \label{modegam}
		\gamma^{\mu\nu}(x,\eta)=\int \frac{d^3k}{\left(2\pi\right)^3} \left(a_{k} u^{\mu\nu}_{k}(\eta)+ a_{-k}^{\dagger} u^{*\mu\nu}_{k}(\eta) \right)e^{ik.x}
	\end{equation}
	with 
	\begin{align}\label{modegravi1}
		&u^{\mu\nu}(k, \eta) = \frac{z^{\mu}z^{\nu}}{\sqrt{2k^3}}[e^{ik\eta}(1-i k\eta)C + e^{-ik\eta}(1+i k\eta)D]
	\end{align}
	where $z^{\mu},z^{\nu}$ are polarization tensors and $C,D$ are some complex numbers satisfying $|C|^2-|D|^2 = 1$
	\begin{equation}\label{albttld}
	C= \cosh\left({\tilde \alpha}\right),~~~~	D= - i e^{i{\tilde \beta}} \sinh\left({\tilde \alpha}\right).
	\end{equation}
	The $| {\tilde \alpha}, {\tilde \beta}\rangle$ vacuum is defined by
	\begin{equation}\label{atlbttvc}
		a_{k} | {\tilde \alpha}, {\tilde \beta}\rangle =0.
	\end{equation}
	
	We now use these mode expansions to calculate the correlation function in $\alpha-$vacua. To calculate the cosmological correlation function we are going to use in-in formalism which gives the correlation function
	\begin{equation} \label{ininf}
	\begin{split}
	&\langle \phi^{int}(x_1, t)\phi^{int}(x_2, t)\cdots\phi^{int}(x_n, t) \rangle  \\
	&= 	\frac{\langle \bar{\mathcal{T}}\left(\exp(i\int_{\infty(1+i\epsilon)}^tdt' H^{int}_{I})\right)\, \prod_{i=1}^{n} \phi^{int}(x_i, t)  \, \mathcal{T}\left(\exp(i\int_{-\infty(1+i\epsilon)}^tdt' H^{int}_{I})\right)\rangle}{\langle \bar{\mathcal{T}}\left(\exp(i\int_{\infty(1+i\epsilon)}^tdt' H^{int}_{I})\right)\mathcal{T}\left(\exp(i\int_{-\infty(1+i\epsilon)}^tdt' H^{int}_{I})\right)\rangle}
	\end{split}
	\end{equation}
	where
	\begin{equation}
	\prod_{i=1}^{n} \phi^{int}(x_i, t) = \phi^{int}(x_1, t)\phi^{int}(x_2, t)\cdots\phi^{int}(x_n, t) \, .
\end{equation}	 
In \eqref{ininf} $\phi^{int}(x_i, t)$ are the fields written in interaction picture, and $H^{int}_{I}$ is the interacting Hamiltonian in the interaction picture \footnote{We will drop the subscripts and the superscripts on the fields in the following discussions to avoid clutter of notations.} 

In the following sub-sections we will use \eqref{ininf} to compute two and three point functions involving scalar and tensor fields. For that we will transform the time coordinate to conformal time variable $\eta$. Once we know the explicit form of $H^{int}_{I}$ relevant for the particular correlation function, the RHS of \eqref{ininf} will be determined perturbatively by bringing down factors of $H^{int}_{I}$ from the exponential. Next, one needs to substitute mode expansions for the fields given in \eqref{modex} and \eqref{modegam}. Finally, one needs to use the desired vacuum given in \eqref{atlbttvc}. Once all these are taken care of, ultimately, we will be left with an integration over the conformal time $\eta$ along a chosen contour from $\eta = -\infty$ to $\eta = 0$ (see \cite{Maldacena:2011nz}). In the following sections we will look into various cases and will perform this $\eta$-integration. 

These correlation functions can be interpreted as correlation functions of some appropriate operators in three-dimensional Euclidean CFT, for details see \cite{Maldacena:2011nz}. For example, a conformally coupled scalar field (denoted by $\varphi$) with mass $m^2 = 2 H^2$ in $dS_4$, will correspond to an operator $O_{\Delta}$ with scaling dimension $\Delta =2$ in $CFT_3$. Similarly, for a massless scalar field $\phi$ in $dS_4$, we will have $O_{\Delta}$ with $\Delta =3$. Also, for a massless vector field $A_\mu$ in $dS_4$, we will get a conserved spin one current $J_i$ (with $\Delta = 2$), and a spin two graviton $\gamma_{\mu\nu}$ in $dS_4$ will correspond to a conserved spin two current $T_{ij}$ (with $\Delta = 3$) in $CFT_3$ \footnote{It should be emphasized that we are not assuming any kind of dS/CFT correspondence here. These statements follow just from the fact that asymptotic isometry of $dS_4$ (as $\eta \rightarrow 0$)  is same as the symmetry group of $3$-dimensional Euclidean CFT.}. Performing the calculation on the RHS of \eqref{ininf}, we will obtain correlation functions involving late time values of the fields $\varphi, \, A_{\mu}, \, \gamma_{\mu\nu}$. Next, following the steps outlined in appendix \ref{Shad} we can translate them to correlators involving the operators $O, \, J,\, T$ in CFT. 

	\subsection{Two-point function} 
	
	In this section we calculate the two point function of scalar and graviton field in rigid dS background using in-in formalism.
	\subsubsection{$\langle \varphi \varphi\rangle$}
	In this case the following time integral can be obtained from \eqref{ininf} for $\langle \varphi \varphi\rangle$. 
	\small
	\begin{align}
		k^2_1 \, \Im\bigg[\int_{-\infty}^0\frac{d\eta}{\eta^2} \bigg( &f_{k_1}(A,B)\bar{u}_{k_1}(\eta)[f_{k_2}(A,B)\bar{u}_{k_2}(\eta)-\bar{f}_{k_2}(A,B)u_{k_2}(\eta)]\notag\\&+f_{k_2}(A,B)\bar{u}_{k_2}(\eta)[f_{k_1}(A,B)\bar{u}_{k_1}(\eta)-\bar{f}_{k_1}(A,B)u_{k_1}(\eta)]-\text{c.c.}\bigg)\bigg]_{k^i_2 \to -k^i_1}
	\end{align}
	\small
	where
	\begin{align}
		f_k(A,B) = \frac{1}{\sqrt{k^3}}(A+B) \label{varfun}
	\end{align}
	and $u$ is defined in \eqref{me}. The ${\bar u}$ is complex conjugate of $u.$
	The time integral after the shadow transform, discussed in appendix \ref{Shad} ,  gives the following correlator
	\begin{align}
		\langle O(k_2)O(-k_2)\rangle = \frac{k^3_2}{(A+B)(\bar{A}+\bar{B})(|A|^2-|B|^2)} =\frac{k^3_2}{(A+B)(\bar{A}+\bar{B})}\label{2pt}
	\end{align}
	where in the last line we have used  $|A|^2-|B|^2=1.$ 
	Now using \eqref{absca} we have 
	\begin{equation}
		\langle O(k_2)O(-k_2)\rangle =  \frac{k^3_2}{(\cosh 2\alpha + \sin\beta \sinh 2\alpha)}\label{2ptab}
	\end{equation}
	and when $\beta=0$ we get
	\begin{equation}
		\langle O(k_2)O(-k_2)\rangle = \frac{1} {\cosh 2\alpha}k_2^3\label{2pta1}.
	\end{equation}
	We now compute two point function of graviton.
	\subsubsection{$\langle TT\rangle$}
	We recalculate $\langle TT\rangle$ in a general vacuum using the mode expansion for the graviton given by (\ref{modegravi1}) and  using the interaction
	\begin{align}
		H_{int} &= \int \sqrt{-g}\, d^4x ~ R = \frac{1}{4}\int \frac{1}{\eta^2} d\eta k^2 d^3{k} \gamma_{\mu\nu}(k)\gamma^{\mu\nu}(-k)
	\end{align}
	in (\ref{ininf}). The time integral one obtains is given by
	\begin{align}
		\frac{1}{4}(z_1.z_2)^2k^2_1\Im\bigg[\int_{-\infty}^0\frac{d\eta}{\eta^2} \bigg( &f_{k_1}(C, D)\bar{\gamma}_{k_1}(\eta)[f_{k_2}(C, D)\bar{\gamma}_{k_2}(\eta)-\bar{f}_{k_2}(C, D)\gamma_{k_2}(\eta)]\nonumber\\
		+&f_{k_2}(C, D)\bar{\gamma}_{k_2}(\eta)[f_{k_1}(C, D)\bar{\gamma}_{k_1}(\eta)-\bar{f}_{k_1}(C, D)\gamma_{k_1}(\eta)]-\textbf{c.c.}\bigg)\bigg]_{k^i_2 \to -k^i_1}
	\end{align}
	where 
	\begin{equation}\label{gamma}
	\gamma_{k}(\eta)=\frac{1}{\sqrt{2k^3}}[e^{ik\eta}(1-i k\eta)C + e^{-ik\eta}(1+i k\eta)D]
	\end{equation}
	 and ${\bar \gamma}$ is just the comeplx conjugate of $\gamma$.
	which when computed, after the shadow transform, leads to 
	\begin{align}\label{tt2pt}
		\langle T(z_1, k_1)T(z_2, k_2)\rangle &= (z_1\cdot z_2)^2\frac{k^3_1}{(C+D)(\bar{C}+\bar{D})(|C|^2-|D|^2)}\nonumber\\
		&=(z_1\cdot z_2)^2\frac{k^3_1}{(C+D)(\bar{C}+\bar{D})}.
	\end{align}
	Now using \eqref{albttld} we get
	\begin{equation}
		\langle T(z_1, k_1)T(z_2, k_2)\rangle = (z_1\cdot z_2)^2\frac{1}{ (\cosh 2\tilde{\alpha}+\sin \tilde{\beta} \sinh 2\tilde{\alpha})} k_2^3\label{2ptabt1}
	\end{equation}
	and when ${\tilde \beta}=0$ we get
	\begin{equation}
		\langle T(z_1, k_1)T(z_2, k_2)\rangle =  (z_1\cdot z_2)^2\frac{1}{\cosh 2{\tilde \alpha}} k_2^3 \label{2pta1b1}.
	\end{equation}

	\subsection{Three-point function}\label{thptall}
	We now turn our attention to calculating the three-point functions. We start with the simplest case of scalars and then move on to spinning fields.
	\subsubsection{$\langle OOO\rangle$}
	We recalculate $\langle OOO\rangle$ in a general vacua characterized
	by the mode expansion (\ref{modex}). Using the mode expansion in (\ref{ininf}), with 
	\begin{align}
		H_{int} = \int d^{4}x\sqrt{-g}~\phi^3
	\end{align}
	we obtain the following time integral
	\begin{align}
		6\Im\bigg[&\int_{-\infty}^0\frac{d\eta}{\eta^4} \bigg([-f_{k_3}(A, B)\bar{u}_{k_3}(\eta)+\bar{f}_{k_3}(A, B)u_{k_3}(\eta)]f_{k_1}(A, B)f_{k_2}(A, B)\bar{u}_{k_1}(\eta)\bar{u}_{k_2}(\eta)\notag\\&+[-f_{k_2}(A, B)\bar{u}_{k_2}(\eta)+\bar{f}_{k_2}(A, B)u_{k_2}(\eta)f_{k_1}(A, B)\bar{f}_{k_3}(A, B)\bar{u}_{k_1}(\eta)u_{k_3}(\eta)+[-f_{k_1}(A, B)\bar{u}_{k_1}(\eta)\notag\\&+\bar{f}_{k_1}(A, B)u_{k_1}(\eta)]\bar{f}_{k_2}(A, B)\bar{f}_{k_3}(A, B)u_{k_2}(\eta)u_{k_3}(\eta)\bigg)\bigg]
	\end{align}
	where $f_k(A,B)$ and $u_k(\eta)$ have been defined previously in (\ref{me}) and eq.\eqref{varfun}. The integral once computed, after the shadow transform, leads to
	\begin{align}
		\langle OOO\rangle = a R_1 + b \left(R_2 +  R_3 +  R_4 \right)\label{cftans}
	\end{align}
	where
	\begin{align}\label{cftans11}
		&a = \frac{1}{2\mathcal{N}^3(A, B)}[(2A^2+3 AB+3B^2)(\bar{A}^2+\bar{B}^2+\bar{A}\bar{B})+(A-B)\bar{B}(|A|^2+|B|^2+A\bar{B})] \nonumber\\
		&b= \frac{1}{2\mathcal{N}^3(A, B)}[(A^2+6AB+ B^2)\bar{A}\bar{B}+A\bar{B}^2(-A+B)+(A-B)B\bar{A}^2]\nonumber\\
		&\mathcal{N}(A,B)=\left(A+B\right)\left(\bar{A}+\bar{B}\right)
	\end{align}
	and
	\begin{align}\label{scal3pt1}
		&R_1 = R(k_1, k_2, k_3) \quad R_2 = R(k_1, k_2, -k_3) \quad R_3 = R(k_1, -k_2, k_3) \quad R_4 =R(-k_1, k_2, k_3) \end{align}
	with
	\begin{align}\label{scal3pt2}
		R(k_1, k_2, k_3)  =  -\frac{4}{9}\sum_a k^3_a-\frac{1}{3}\sum_{a\neq b} k^2_a k_b+\frac{1}{3}k_1k_2k_3 + 3\sum_a k^3_a\log E
	\end{align}
	where $E=k_1+k_2+k_3.$
	Notice the bulk computation automatically forces exchange symmetry. Also, note that $a, b$ is real i.e. under $A \leftrightarrow \bar{A}, B \leftrightarrow \bar{B}$ the coefficients $a, b$ are invariant meaning $a^{*} = a, b^* = b$.
	Using the parameterization as in \eqref{absca} we have \begin{align}
		\langle OOO\rangle =a R_1 +  b \left(R_2 +  R_3 +  R_4 \right)\label{cftansab}
	\end{align}
	where 
	\begin{align}\label{abres}
		&a= \frac{5+3\cosh 4\alpha-6\cos 2\beta \sinh^2 2\alpha+6 \sin\beta \sinh 4\alpha}{8(\cosh 2\alpha + \sin\beta \sinh 2\alpha)^3}\nonumber\\
		& b= \frac{[\cosh 2\alpha \sin\beta +(3+\cos 2\beta)\cosh\alpha \sinh\alpha]\sinh 2\alpha}{2(\cosh 2\alpha + \sin\beta \sinh 2\alpha)^3}
	\end{align}
	For the special case of $\beta=0$ we have 
		\begin{align}\label{abres1}
	&a= \frac{1}{ \cosh^3 2\alpha  }\nonumber\\
	& b= \frac{ \sinh^2\alpha}{ \cosh^3 2\alpha  }
	\end{align}
	which gives
	\begin{align}
		\langle OOO\rangle =  \frac{1}{\cosh^3 2\alpha}[R_1 +  \sinh^2 2\alpha\left(R_2 +  R_3 +  R_4 \right)]\label{cftansab0}.
	\end{align}
	The results in \eqref{cftansab0} is same as that appears in \cite{Shukla:2016bnu}.
	\subsection*{Spinning correlator}
	For this case we need to define the vacuum suitably. 
	We have characterised the vacuum for scalar mode in \eqref{albtvc} and for tensor mode in \eqref{atlbttvc}. In general when we consider correlation function involving both graviton and scalar we can conside most general vacuum defined of the form $|\alpha,\beta, {\tilde \alpha}, {\tilde \beta} \rangle.$ However to start with let us take simpler situation when we have 
	$A=C, B=D$ which implies $\alpha={\tilde \alpha}$ and $\beta={\tilde \beta}$.
	\subsubsection{$\langle TOO\rangle$}
	To compute two scalars and one graviton amplitude we need to consider 
	\begin{align}
		H_{int} = \int d^4x \sqrt{-g}~g^{\mu\nu}\partial_{\mu}\phi\partial_{\nu}\phi    \end{align}
	Here we just provide the final results, the computational details of which are provided in Appendix \ref{gtoob}.
	The correlation function is given by
	\begin{align}
		\langle TO_3O_3\rangle = a\langle TO_3O_3\rangle_{R_1} +b\left(\langle TO_3O_3\rangle_{R_2}+\langle TO_3O_3\rangle_{R_3}+\langle TO_3O_3\rangle_{R_4}\right)\label{TOOg1}  
	\end{align}
	where the form of $a,b$ is same as that appears in \eqref{cftans11}
	and 
	\small
	\begin{align}
		&\langle TO_3O_3\rangle_{R_1} =\left[k_1+k_2+k_3-\frac{k_1k_2+k_2k_3+k_3k_1}{k_1+k_2+k_3}-\frac{k_1k_2k_3}{(k_1+k_2+k_3)^2}\right] (z_1.k_2)^2\nonumber\\
		&\langle TO_3O_3\rangle_{R_2} = \left[-k_1+k_2+k_3-\frac{-k_1k_2+k_2k_3-k_3k_1}{-k_1+k_2+k_3}+\frac{k_1k_2k_3}{(-k_1+k_2+k_3)^2}\right](z_1.k_2)^2\nonumber\\
		&\langle TO_3O_3\rangle_{R_3} =  \left[k_1-k_2+k_3-\frac{-k_1k_2-k_2k_3+k_3k_1}{k_1-k_2+k_3}+\frac{k_1k_2k_3}{(k_1-k_2+k_3)^2}\right](z_1.k_2)^2\nonumber\\
		&\langle TO_3O_3\rangle_{R_4} = \left[k_1+k_2-k_3-\frac{k_1k_2-k_2k_3-k_3k_1}{k_1+k_2-k_3}+\frac{k_1k_2k_3}{(k_1+k_2-k_3)^2}\right](z_1.k_2)^2\label{corr1a}
	\end{align}

	\subsubsection{ $\langle TTO_3\rangle$}
	To calculate two gravitons and one scalar amplitude
	we consider the following interaction term
	\begin{equation}\label{inttto1}
		H_{int}=\int d^4x~\sqrt{-g}\varphi {\mathcal W}_{\rho\sigma\alpha \beta} {\mathcal W}_{\rho \sigma \alpha \beta}
	\end{equation} 
	Here we write down the final result, details of the computation are provided in appendix \ref{TTobg}. 
	The correlation function is given by
	\begin{align}\label{tto3riab}
		\langle TTO_3\rangle_{\alpha} = a\langle TTO_3\rangle_{R_1}+b\left(\langle TTO_3\rangle_{R_4}+\langle TTO_3\rangle_{R_2} + \langle TTO_3\rangle_{R_3}\right).
	\end{align}
	where the form of $a,b$ is same as that appears in \eqref{cftans11}.
	The explicit expressions for $\langle TTO_3 \rangle_{R_i}$ are complicated. They are best written in spinor helicity variables. We give their explicit forms in section \ref{tto3ab}.
	\subsubsection{$\langle TTT\rangle$}\label{TTTgb11}
	The three graviton amplitude can get contribution from two different sources, the Einstein Hilbert term and the $Weyl^3$  term. Let us start with the Weyl tensor contribution.
	\subsubsection*{$Weyl^3$ contribution}
	To calculate the three graviton amplitude we need to consider the following interaction 
	\begin{align}
		S^{(3)}_{\gamma, {\mathcal W}^3} = \int d^4x\sqrt{-g}~ {\mathcal W}^3 \label{hI}.
	\end{align}
 where ${\mathcal W}_{abcd}$ is the Weyl tensor. 
	We provide computational details in appendix \ref{TTTgb1}.
	The correlation functions are given by
	\begin{align}
		\langle TTT\rangle_{{\mathcal W}^3, \alpha} = a\langle TTT\rangle_{{\mathcal W}^3, 1}+b(\langle TTT\rangle_{{\mathcal W}^3, 2}+\langle TTT\rangle_{{\mathcal W}^3, 3}+\langle TTT\rangle_{{\mathcal W}^3, 4})\label{hgtttgr}
	\end{align}
	where $a, b$ are  same as that appears in \eqref{cftans11}.
	We also have 
	\begin{align}
		\langle TTT\rangle_{{\mathcal W}^3, i} = \langle TTT\rangle_{h, R_i}
	\end{align}
	where $\langle TTT\rangle_{h, R_i}$ appear in (\ref{homttt}), \eqref{homtttsh}. 
	\subsubsection*{Two-derivative interaction the Einstein-Hilbert contribution}
	Consider now the interaction \footnote{Before we put our paper on arXiv, \cite{Kanno:2022mkx} appeared which  computes this correlation function and discusses its phenomenological implications.}
	\begin{align}
		S^{(3)}_{\gamma, EG} = \int d^4x\sqrt{-g}~ R
	\end{align}
	Again details are provided  in Appendix \ref{TTTgb1}.
	Correlation function in this case is given by
	\begin{align}
		\langle TTT\rangle_{EG, \alpha} = a\langle TTT\rangle_{EG, 1}+b(\langle TTT\rangle_{EG, 2}+\langle TTT\rangle_{EG, 3}+\langle TTT\rangle_{EG, 4})\label{Tnhttg1}
	\end{align}
	where $a, b$ are precisely given again by \eqref{cftans11} and 
	\begin{align}
		\langle TTT\rangle_{EG, i} = \langle TTT\rangle_{nh, R_i}\quad i = 1, 2, 3, 4
	\end{align}
	where $\langle TTT\rangle_{nh, R_i}$'s appear in (\ref{othhno}).
	\subsection{Spinning correlator in general vacuum}\label{mxdgnl1}
	In this subsection we consider more spinning correlaltor of the form $\langle T OO\rangle$ and $\langle TT O\rangle.$ In earlier section we consider this correlation function for the special case when $A=C, B=D$ which implies $\alpha={\tilde \alpha}$ and $\beta={\tilde \beta}$. We now consider the case when $A\neq C, B\neq D$ which implies $\alpha\neq {\tilde \alpha}$ and $\beta \neq {\tilde \beta}.$ For this case the general vacuum is denoted by $| \alpha,\beta,{\tilde \alpha},{\tilde \beta}\rangle.$
	For this case as well calculation is straightforward and is given in Appendix \ref{gtoob} and \ref{TTobg}.
	The $\langle T OO\rangle$ is then obtained to be
	\begin{align}
		\langle TO_3O_3\rangle = c_1\langle TO_3O_3\rangle_{R_1} +c_2\langle TO_3O_3\rangle_{R_2}+c_3 \left(\langle TO_3O_3\rangle_{R_3}+\langle TO_3O_3\rangle_{R_4}\right)\label{TOOg11}
	\end{align}
	where $c_i$ are given in \eqref{abc}. Let us note that as compared to \eqref{TOOg1} where the correlator was parameterised by two parameters, here  we have extra parameter. The Ward-Takahashi(WT) identity is given by
	\begin{equation}
		k^{\mu}_1z^{\nu}\langle T_{\mu\nu} O_3 O_3\rangle =(c_1+c_2)(z_1.k_2) (k^3_2-k^3_3).
	\end{equation}
	For the case of $\langle TTO_3\rangle$ we have
	\begin{align}\label{tto3gab}
		\langle TTO_3\rangle_{\alpha} = d_1\langle TTO_3\rangle_{R_1}+d_2\langle TTO_3\rangle_{R_4}+ d_3(\langle TTO_3\rangle_{R_2} + \langle TTO_3\rangle_{R_3})
	\end{align}
	where $d_i$ are given in \eqref{di123}. Again, let us note that as compared to \eqref{tto3riab} where the correlator was parameterised by two parameters, here  we have extra parameter and is thus parameterised by three parameters.

	\section{Spinning CFT correlators using weight shifting and spin raising operator}\label{spinCFT}
	In this section, we discuss how to understand the results obtained in the previous section from a CFT perspective. Let us start our discussion with a two-point function. 
	
	We shall be using the notation $k\equiv |\vec{k}|$ in the following. We also choose to work with null and transverse polarization tensor such that $z_i.k_i=0, z_i^2=0.$
	\subsection{Two-point function}
	Consider the following two-point functions of stress tensor and scalar operator $O_3$
	\begin{align}
		&\langle T(z_1, k_1)T(z_2, k_2)\rangle = C_T (z_1.z_2)^2 k^3_1\\
		&\langle O_{3}(k_1)O_{3}(k_2)\rangle = C_{O} k^3_3
	\end{align}
	Comparing this result with general $\alpha-$vacuum answer \eqref{tt2pt} and \eqref{2pt1} one can obtain $C_{O},C_T.$
	
	\subsection{Three-point function}
	We now turn our attention to computing the three-point function. We start with the scalar three-point function. The momentum space result for the three-point function is much richer than the position space counterpart. For scalar three-point function, there are four different solutions to conformal ward identity \cite{Bzowski:2013sza}. If we impose consistency with OPE limit, only one particular combination of four solutions survives, which matches with the cosmological correlation function computed in the Bunch-Davies (BD) vacuum and is also can be thought of as a Fourier transform of position space answer. As will be shown in this section, the cosmological correlation function in general $\alpha-$vacua requires us to consider even the solutions which are not consistent with OPE limit. As will be shown below, this is true even for correlation functions with spinning operators.

	\subsubsection{$\langle O_3O_3O_3\rangle$}
	For the case of scalar three point function, we have four solutions to conformal ward identity. 
	These are given by \cite{Bzowski:2013sza}
	\begin{align}
		&R_1 = A(k_1, k_2, k_3) \quad R_2 = A(k_1, k_2, -k_3) \quad R_3 = A(k_1, -k_2, k_3) \quad R_4 = A(-k_1, k_2, k_3) 
	\end{align}
	where
	\begin{align}
		A(k_1, k_2, k_3)  =  -\frac{4}{9}\sum_a k^3_a-\frac{1}{3}\sum_{a\neq b} k^2_a k_b+\frac{1}{3}k_1k_2k_3 + 3\sum_a k^3_a\log E
	\end{align}
	where $E=k_1+k_2+k_3.$
	The most general answer to three point function is given by
	\begin{align}
		\langle O_3(k_1)O_3(k_2)O_3(k_3)\rangle = c_1 R_1+c_2 R_2+ c_3 R_3+c_4 R_4\label{gvac}.
	\end{align}
	Consistency with OPE limit sets $c_2=c_3=c_3=0$ and gives the Bunch-Davies (BD) vacuum answer
	\begin{equation}
		\langle O_3(k_1)O_3(k_2)O_3(k_3)\rangle_{BD} =   -\frac{4}{9}\sum_a k^3_a-\frac{1}{3}\sum_{a\neq b} k^2_a k_b+\frac{1}{3}k_1k_2k_3 + 3\sum_a k^3_a\log E.
	\end{equation}
	However, if we don't demand consistency with OPE, we obtain the most general answer given in \eqref{gvac}. The exchange symmetry $1 \leftrightarrow 2  \leftrightarrow 3$ requires $c_2=c_3=c_4$ which gives
	\begin{align}
		\langle O_3(k_1)O_3(k_2)O_3(k_3)\rangle = c_1 R_1+c_2(R_2+R_3+R_4).
	\end{align}
	Let us note that, $c_1$ and $c_2$ are real numbers because the correlator is real. Comparing with general $\alpha-$vacua result \eqref{cftans},\eqref{cftans11}  we obtain values of parameter $c_1,c_2$
	\begin{align}\label{c12a}
		&c_1 = a, c_2= b
	\end{align}
	
	Having discussed the scalar three-point function case, let us consider the spinning correlator. The story here is much richer than the scalar case. Let us start with the simplest of the spinning correlation function $\langle TO_3O_3\rangle$.
	
	\subsubsection{$\langle TO_3O_3\rangle$}
	The easiest way to get to the result is to use the spin and dimension raising operator starting from $ \langle O_3O_3O_3\rangle$. The spin and dimension raising operator is reviewed in the appendix \ref{spinrai}.
	The correlator $\langle TO_3O_3\rangle$ can be obtained by
	\begin{align}
		\langle TO_3O_3\rangle = D_{13}D_{12}{\mathcal W}^{++}_{23}\langle O_3O_3O_3\rangle
	\end{align}
	which in the momentum space is given by
	\begin{align}\label{avac1}
		\langle TO_3O_3\rangle_{\alpha} = a\langle TO_3O_3\rangle_{R_1}+b(\langle TO_3O_3\rangle_{R_2}+\langle TO_3O_3\rangle_{R_3}+\langle TO_3O_3\rangle_{R_4})
	\end{align}
	where $a,b$ are precisely as given in \eqref{cftans11}
	and $\langle TO_3O_3\rangle_{R_i}$ is same as that appears in \eqref{corr1a}. 
	This also has satisfies Ward identity given by
	\begin{align}
		k^{\mu}_1z^{\nu}\langle T_{\mu\nu} O_3 O_3\rangle &=(a+b)(z_1.k_2) (k^3_2-k^3_3)\notag\\
		&= (z_1.k_2) (\langle O_3(k_2)O_3(-k_2)\rangle-\langle O_2(k_3)O_2(-k_3)\rangle)
	\end{align}
	where we have identified $a+b = C_{\Delta = 3}$.
	We would like to understand the results from the perspective of solutions of conformal ward identity as we did for the case of scalar correlator in the previous subsection. We do this in the next section.

	\subsubsection{$\langle JJO_3\rangle$}
	Let us now consider $\langle JJO_3\rangle$.
	The correlator $\langle JJO_3\rangle$ may be computed from $\langle O_3O_3O_3\rangle$ using spin raising and weight shifting operators as follows
	\begin{align}
		\langle JJO_3\rangle = k_1 k_2 H_{12}{\mathcal W}^{--}_{12}\langle O_3O_3O_3\rangle
	\end{align}
	which in momentum space is  given by
	\begin{align}\label{fnlansj30}
		\langle JJO_3\rangle = a\langle JJO_3\rangle_{R_1}+b(\langle JJO_3\rangle_{R_2}+\langle JJO_3\rangle_{R_3}+\langle JJO_3\rangle_{R_4})
	\end{align}
	where 
	\begin{align}
		&\langle JJO_3\rangle_{R_1} = \frac{12 (k_1+k_2+2k_3)}{(k_1+k_2+k_3)^2}[(k_1+k_2+k_3)(k_1+k_2-k_3)(z_1.z_2)+2z_1.k_2z_2.k_1]\notag\\
		&\langle JJO_3\rangle_{R_2} = \frac{12 (-k_1+k_2+2k_3)}{(k_2+k_3-k_1)^2} [(k_1-k_2-k_3)(k_1+k_3-k_2)(z_1.z_2)+2z_1.k_2 z_2.k_1] \notag\\
		&\langle JJO_3\rangle_{R_3} = \frac{12(k_1-k_2+2k_3)}{(k_1-k_2+k_3)^2}[(k_1-k_2+k_3)(k_1-k_3-k_2)(z_1.z_2)+2z_1.k_2 z_2.k_1]\notag\\
		&\langle JJO_3\rangle_{R_4} = \frac{12(k_1+k_2-2k_3)}{(k_1+k_2-k_3)^2}[(k_1+k_2+k_3)(k_1+k_2-k_3)(z_1.z_2)+2z_1.k_2 z_2.k_1]  \label{basisJJO}
	\end{align}
	Again demanding consistency with OPE limit would give
	\begin{align}\label{fnlansj301}
		\langle JJO_3\rangle = \langle JJO_3\rangle_{R_1}
	\end{align}
	which is precisely the correlation function obtained in Bunch-Davies vacuum.
	In spinor helicity variables, results looks much simpler and takes the following form
	\begin{align}
		&\langle J^{-}J^{-}O_3\rangle_{R_1} = \frac{12 (k_1+k_2+2k_3)}{(k_1+k_2+k_3)^2}\langle 12\rangle^2 \quad \langle J^{-}J^{-}O_3\rangle_{R_4} = \frac{12 (k_1+k_2-2k_3)}{(k_1+k_2-k_3)^2}\langle 12\rangle^2\\
		&\langle J^{-}J^{-}O_3\rangle_{R_2} = \langle J^{-}J^{-}O_3\rangle_{R_3} = 0
	\end{align}
	\begin{align}
		&\langle J^{-}J^{+}O_3\rangle_{R_1} =  \langle J^{-}J^{+}O_3\rangle_{R_4} = 0\notag\\
		&\langle J^{-}J^{+}O_3\rangle_{R_2} = \frac{12 (-k_1+k_2+2k_3)}{E^2(k_2+k_3-k_1)^2}\langle 31\rangle^2\langle\bar{2}\bar{3}\rangle^2 \quad \langle J^{-}J^{+}O_3\rangle_{R_3} = \frac{12(k_1-k_2+2k_3)}{E^2(k_1-k_2+k_3)^2}\langle 31\rangle^2\langle\bar{2}\bar{3}\rangle^2
	\end{align}
	and its complex conjugates. 
	One can check that this result is consistent with calculation in dS space if we consider 
	\begin{equation}
		\int \varphi F_{\mu\nu}F^{\mu\nu}. 
	\end{equation}
	
	The relation between the correlators in general $\alpha-$vacua and BD vacuum can be written as 
	\begin{align}
		\langle JJO_3\rangle_{\alpha}-(a+b) \langle JJO_3\rangle_{BD} = -b[\langle JJO_3\rangle_{R_1} - \langle JJO_3\rangle_{R_2} - \langle JJO_3\rangle_{R_3} - \langle JJO_3\rangle_{R_3}].
	\end{align}

	\subsection{$\langle TTO_3\rangle$}\label{tto3ab}
	Similarly, correlator $\langle TTO_3\rangle$ may be computed from $\langle O_3O_3O_3\rangle$ using
	\begin{align}
		\langle TTO_3\rangle = k^3_1 k^3_2 H^2_{12}{\mathcal W}^{--}_{12}\langle O_3O_3O_3\rangle
	\end{align}
	which again in momentum space becomes
	\begin{align}
		\langle TTO_3\rangle = a\langle TTO_3\rangle_{R_1}+b(\langle TTO_3\rangle_{R_2}+\langle TTO_3\rangle_{R_3}+\langle TTO_3\rangle_{R_4})
	\end{align}
	In momentum space $\langle TTO_3\rangle$ is complicated, however in spinor helcity variables they take simple form and are given by
	\begin{align}
		&\langle T^{-}T^{-}O_3\rangle_{R_1} = \frac{48 k_1k_2(k_1+k_2+4k_3)}{(k_1+k_2+k_3)^4}\langle 12\rangle^4 \quad \langle T^{-}T^{-}O_3\rangle_{R_4} = \frac{48 k_1k_2(k_1+k_2-4k_3)}{(k_1+k_2-k_3)^4}\langle 12\rangle^4\nonumber\\
		&\langle T^{-}T^{-}O_3\rangle_{R_2} = \langle T^{-}T^{-}O_3\rangle_{R_3} = 0.\label{avac2}
	\end{align}
	For the mixed helicity we have
	\begin{align}
		&\langle T^{-}T^{+}O_3\rangle_{R_1} =  \langle T^{-}T^{+}O_3\rangle_{R_4} = 0\nonumber\\
		&\langle T^{-}T^{+}O_3\rangle_{R_2} = \frac{48k_1k_2 (-k_1+k_2+4k_3)}{E^4 (k_2+k_3-k_1)^4}\langle 31\rangle^4\langle \bar{2}\bar{3}\rangle^4 \quad \langle T^{-}T^{+}O_3\rangle_{R_3} = \frac{48k_1k_2(k_1-k_2+4k_3)}{E^4(k_1-k_2+k_3)^4}\langle 31\rangle^4\langle \bar{2}\bar{3}\rangle^4.\label{avac3}
	\end{align}
	In general vacua, we have
	\begin{align}
		\langle TTO_3\rangle_{\alpha}-(a+b) \langle TTO_3\rangle_{BD} = -b[\langle TTO_3\rangle_{R_1} - \langle TTO_3\rangle_{R_2} - \langle TTO_3\rangle_{R_3} - \langle TTO_3\rangle_{R_4}]
	\end{align}
	In spinor-helicity variables, the independent components are
	\begin{align}
		& \langle T^{-}T^{-}O_3\rangle_{\alpha}-(a+b) \langle T^{-}T^{-}O_3\rangle_{BD} = -b (\langle T^{-}T^{-}O_3\rangle_{R_1}-\langle T^{-}T^{-}O_3\rangle_{R_4})\\
		& \langle T^{-}T^{+}O_3\rangle_{\alpha}-(a+b) \langle T^{-}T^{+}O_3\rangle_{BD} = -b (\langle T^{-}T^{+}O_3\rangle_{R_2}+\langle T^{-}T^{+}O_3\rangle_{R_3})
	\end{align}
	
	\subsection{$\langle TTT\rangle $}\label{TTTh}
	Let us now turn our attention to $\langle TTT\rangle.$ The ward takahashi identity is given by
	\begin{align}\label{Twtnjh}
		k_1^{\mu}z_1^{\nu}\langle T_{\mu\nu}T(z_2, k_2)T(z_3, k_3)\rangle =&-\left(z_{1}.k_2\right) z_{2}^{i} z_{2}^{j}\langle T_{\vec{k}_{2}+\vec{k}_{1}}^{i j} T_{\vec{k}_{3}}\rangle+2\left(z_1.z_2\right) z_{2}^{j} k_{2}^{i}\langle T_{\vec{k}_{2}+\vec{k}_{1}}^{i j} T_{\vec{k}_{3}}\rangle \notag\\
		&-\left(z_{1} \cdot k_{3}\right) z_{3}^{i} z_{3}^{j}\langle T_{\vec{k}_{2}} T_{\vec{k}_{3}+\vec{k}_{1}}^{i j}\rangle+2(z_1.z_3) z_{3}^{j} k_{3}^{i}\langle T_{\vec{k}_{2}} T_{\vec{k}_{3}+\vec{k}_{1}}^{i j}\rangle\notag\\
		&+(z_2.k_1)z_{1}^{i} z_{2}^{j}\langle T_{\vec{k}_{2}+\vec{k}_{1}}^{i j} T_{\vec{k}_{3}}\rangle +(z_1.z_2) k_{1}^{i} z_{2}^{j}\langle T_{\vec{k}_{2}+\vec{k}_{1}}^{i j} T_{\vec{k}_{3}}\rangle\notag\\
		&+(z_3.k_1) z_{1}^{i} z_{3}^{j}\langle T_{\vec{k}_{2}} T_{\vec{k}_{1}+\vec{k}_{3}}^{i j}\rangle+(z_1.z_3) k_{1}^{i} z_{3}^{j}\langle T_{\vec{k}_{2}} T_{\vec{k}_{1}+\vec{k}_{3}}^{i j}\rangle.
	\end{align}
	So the solution for $\langle TTT\rangle$ is given by \cite{Jain:2021vrv}
	\begin{equation}
		\langle TTT\rangle= \langle TTT\rangle_{h}+\langle TTT\rangle_{nh}.  
	\end{equation}
	\subsubsection{Homogenous part of $\langle TTT\rangle$}
	The homogenous part of $\langle TTT\rangle$ may be computed from $\langle O_3O_3O_3\rangle$ using spin-raising and weight shifting operators as follows \cite{Baumann:2019oyu,Baumann:2020dch},
	\begin{align}
		\langle TTT\rangle = k^3_3 k^3_2 ({\mathcal W}^{--}_{23})^2(S^{++}_{23})^2D_{13}D_{12}\langle O_3O_3O_3\rangle\label{spdrttt}
	\end{align}
	which  gives
	\begin{align}\label{hansfull}
		\langle TTT\rangle = a\langle TTT\rangle_{h,R_1}+b(\langle TTT\rangle_{h,R_2}+\langle TTT\rangle_{h,R_3}+\langle TTT\rangle_{h,R_4})
	\end{align}
	In the spinor helicity variables, we have
	\begin{align}
		&\langle T^{-}T^{-}T^{-}\rangle_{h,R_1} = \frac{k_1k_2k_3}{(k_1+k_2+k_3)^6}\langle 12\rangle^2\langle 23\rangle^2\langle 31\rangle^2\\
		&\langle T^{-}T^{-}T^{-}\rangle_{h,R_2} = \langle T^{-}T^{-}T^{-}\rangle_{h,R_3} = \langle T^{-}T^{-}T^{-}\rangle_{h,R_4} = 0
	\end{align}
	\begin{align}
		&\langle T^{-}T^{-}T^{+}\rangle_{h,R_4} = \frac{k_1k_2k_3}{(k_1+k_2-k_3)^6}\langle 12\rangle^2\langle 2\bar{3}\rangle^2\langle \bar{3}1\rangle^2\\
		&\langle T^{-}T^{-}T^{+}\rangle_{h,R_2} = \langle T^{-}T^{-}T^{+}\rangle_{h,R_3} = \langle T^{-}T^{-}T^{+}\rangle_{h,R_1} = 0
	\end{align}
	
	\begin{align}
		&\langle T^{+}T^{-}T^{-}\rangle_{h,R_2} = \frac{k_1k_2k_3}{(k_2+k_3-k_1)^6}\langle \bar{1}2\rangle^2\langle 23\rangle^2\langle 3\bar{1}\rangle^2\\
		&\langle T^{+}T^{-}T^{-}\rangle_{h,R_4} = \langle T^{+}T^{-}T^{-}\rangle_{h,R_3} = \langle T^{+}T^{-}T^{-}\rangle_{h,R_1} = 0
	\end{align}
	
	\begin{align}
		&\langle T^{-}T^{+}T^{-}\rangle_{h,R_3} = \frac{k_1k_2k_3}{(k_1+k_3-k_2)^6}\langle 1\bar{2}\rangle^2\langle \bar{2}3\rangle^2\langle 31\rangle^2\\
		&\langle T^{-}T^{+}T^{-}\rangle_{h,R_2} = \langle T^{-}T^{+}T^{-}\rangle_{h,R_4} = \langle T^{-}T^{+}T^{-}\rangle_{h,R_1} = 0
	\end{align}
	The results $\langle TTT\rangle_{h,R_i}$ are precisely same as that appear in \eqref{hgtttgr}. This implies that the homogeneous part is same as that obtained from Weyl term \eqref{hgtttgr} {{${ W}^3$}} in dS space \cite{Jain:2021qcl}.
	Again the following interesting relation between general $\alpha-$vacua and Bunch-Davies vacuum holds,
	\begin{align}
		\langle TTT\rangle_{h, \alpha}-(a+b) \langle TTT\rangle_{h, BD} &=-b(\langle TTT\rangle_{h,R_1} - \langle TTT\rangle_{h,R_2} - \langle TTT\rangle_{h,R_3} - \langle TTT\rangle_{h,R_4}).
	\end{align}
	
	\subsubsection{Non-homogeneous solution}
	The non-homgenous part of $\langle TTT\rangle$ is computed as follows using the weight shifting and spin raising operators \cite{Baumann:2019oyu,Baumann:2020dch}
	\begin{align}
		&\langle T T T\rangle_{A_i}=\mathcal{S}_{31}^{++} \mathcal{S}_{23}^{++} \mathcal{S}_{12}^{++}R_i(k_1, k_2, k_3) \\
		&\langle T T T\rangle_{B_i}=\left(\mathcal{S}_{23}^{++}\right)^{2} \mathcal{W}_{23}^{++} D_{13} D_{12}R_i(k_1, k_2, k_3)+\text { perms. } \\
		&\langle T T T\rangle_{C_i}=D_{13} D_{12}\left(\mathcal{S}_{23}^{++}\right)^{2} \mathcal{W}_{23}^{++}R_i(k_1, k_2, k_3)+\text { perms. }
	\end{align}
	There is also a contact term which is added
	\begin{align}
		&\langle T T T\rangle_{D_i} = (s_{1i}k^3_1+s_{2i}k^3_2+s_{3i}k^3_3)z_1.z_2 z_2.z_3z_3.z_1
	\end{align}
	where $s_i = \pm$ depending on the basis we are working with. The full $\langle TTT\rangle$ is given by
	\begin{align}\label{nhrit}
		\langle TTT\rangle_{nh, R_i} = \frac{1}{4}\langle T T T\rangle_{A_i}-\frac{7}{108}\langle T T T\rangle_{B_i}-\frac{1}{135}\langle TTT\rangle_{C_i} -\frac{36}{5} \langle TTT\rangle_{D_i}
	\end{align}
	where $\langle TTT\rangle_{nh,R_i}$ are given by 
	\begin{align}
		&\langle TTT\rangle_{nh, R_1} = A_1(k_1, k_2, k_3) (z_3.k_1 z_1.z_2 + z_1.k_2 z_2.z_3+z_2.k_3 z_3.z_1)^2\\
		& \sum_{i = 2}^4 \langle TTT\rangle_{nh, R_i} = A_2(k_1, k_2, k_3) (z_3.k_1 z_1.z_2 + z_1.k_2 z_2.z_3+z_2.k_3 z_3.z_1)^2
	\end{align}
	where
	\begin{align}
		& A_2(k_1, k_2, k_3) = A_1(-k_1, k_2, k_3)+ A_1( k_1, -k_2, k_3) + A_1(k_1, k_2, -k_3)\\
		& A_1(k_1, k_2, k_3) = \frac{E^3+(k_1^2+k_2^2+k_3^2)E -2 k_1 k_2 k_3}{2 E^2}
	\end{align}
	The full non-homogeneous answer is given by
	\begin{align}\label{nhansfull}
		\langle TTT\rangle_{nh} = a\langle TTT\rangle_{nh,R_1}+b(\langle TTT\rangle_{nh,R_2}+\langle TTT\rangle_{nh,R_3}+\langle TTT\rangle_{nh,R_4}).
	\end{align}
	Let us note that this is precisely same as that obtained from Einstein-term in dS space, see \eqref{Tnhttg1}.
	For the general vacua, we have
	\begin{align}\label{nhabdch}
		\langle TTT\rangle_{nh, \alpha}-(a+b) \langle TTT\rangle_{nh, BD} &=-b(\langle TTT\rangle_{nh, R_1} - \langle TTT\rangle_{nh, R_2} - \langle TTT\rangle_{nh, R_3} - \langle TTT\rangle_{nh, R_4})
	\end{align}
	
	To summarize this section, we have shown that scalar correlation function in $|\alpha,\beta\rangle$ vacuum can be obtained in three-dimensional CFT by taking a linear combination of four solutions of conformal ward identity of and imposing permutation invariance. For the spinning correlator, we have shown that by acting with spin and dimension raising operators on this scalar correlator we can obtain the correlation function that was obtained in $\alpha-$vacua. However, it is important to note that, spin and dimension raising operators do not reproduce the correlation functions in the generalized vacuum discussed in section \ref{mxdgnl1}.

	\section{Spinning correlator in $\alpha-$vacua by solving conformal ward identity}\label{generalsol}
	The aim of this section is two-fold. First, we want to understand the results previous section about spinning correlator starting from a solution of conformal ward identity and second, we would like to understand the correlation function in a more generalized vacuum $|\alpha, \beta, {\tilde \alpha},{\tilde \beta}\rangle$ that appeared in section \ref{mxdgnl1}.  Let us start our discussion with the simplest example of $\langle T O_3 O_3 \rangle.$
	
	\subsection{$\langle T O_3 O_3 \rangle$}
	Ward identity for $\langle T O_3 O_3 \rangle$ is given by
	\begin{equation}
		\langle k_{1,\mu}T^{\mu\nu}(k_1) O(k_2)O(k_3)\rangle\propto k_{2,\nu} \left(\langle O(k_2)O(-k_2)\rangle-\langle O(k_3)O(-k_3)\rangle\right).  
	\end{equation}
	We can now solve conformal ward identity in spinor-helicity variables, for details see Appendix \ref{sphrep11}.  
	Let us split the solution into homogeneous and non-homogeneous parts as follows
	\begin{align}
		\langle TO_3O_3\rangle = \langle TO_3O_3\rangle_{h} + \langle TO_3O_3\rangle_{nh}.
	\end{align}
	One can show that there are two homogeneous solutions and are given by 
	\begin{align}\label{ohom} 
		&\langle TO_3O_3\rangle_{h_1} = \frac{k_1(k^2_1-k^2_2-4k_2k_3-k^2_3)}{E^2}\left(\frac{\langle 12\rangle\langle 13\rangle}{\langle 23\rangle}\right)^2\notag\\ 
		&\langle TO_3O_3\rangle_{h_2} = \frac{k_1(E-2k_1)^2(-k^2_1+k^2_2-4 k_2k_3+k_3^2)}{(E-2k_3)^2(E-2k_2)^2}\left(\frac{\langle 12\rangle\langle 13\rangle}{\langle 23\rangle}\right)^2 
	\end{align}
	and the non-homogeneous solution is given by
	\begin{align}
		\langle TO_3O_3\rangle_{nh} = \frac{(k_1+k_3-k_1)^2}{k^2_1}\left[k_1+k_2+k_3-\frac{k_1k_2+k_2k_3+k_3k_1}{k_1+k_2+k_3}-\frac{k_1k_2k_3}{(k_1+k_2+k_3)^2}\right]\left(\frac{\langle 12\rangle\langle 13\rangle}{\langle 23\rangle}\right)^2\label{onhom}
	\end{align}
	The most general solution is given by
	\begin{equation}\label{a123}
		\langle TO_3O_3\rangle_{\alpha} = a_1 \langle TO_3O_3\rangle_{nh}+ a_2\langle TO_3O_3\rangle_{h_1}+ a_3\langle TO_3O_3\rangle_{h_2}.  
	\end{equation}
	We would like compare this with \eqref{TOOg11}.
	Consistency with WT identity \footnote{\begin{equation}
			k^{\mu}_1z^{\nu}\langle T_{\mu\nu} O_3 O_3\rangle =(c_1+c_2)(z_1.k_2) (k^3_2-k^3_3).
	\end{equation}} of \eqref{TOOg11} and \eqref{a123} 
	implies
	$a_1=c_1+c_2.$ However the coefficient $a_2,a_3$ in \eqref{a123} remains undetermined. Let us note here that, if we demand consistency with OPE limit we need to set $a_2=a_3=0,$ see \cite{Jain:2021whr} for details. This is consistent with the fact that in BD vacuum, we only have non-homogeneous solution. However, if we relax the consistency with OPE limit we see that both $a_2$ and $a_3$ are non-zero. One can determine both $a_2,a_3$ by comparing \eqref{a123} with \eqref{TOOg11}, \eqref{abc}. One can also identify \footnote{It is interesting to note that   the homogeneous solutions \eqref{h1h2} satisfies MLT test \cite{Jazayeri:2021fvk} even though they are disallowed by OPE consistency condition.} 
	\begin{align}\label{h1h2}
		&\langle TO_3O_3\rangle_{h_1} = 2\left(\langle TO_3O_3\rangle_{R_1}-\langle TO_3O_3\rangle_{R_2}\right)\nonumber\\
		&\langle TO_3O_3\rangle_{h_2} =  -2\left(\langle TO_3O_3\rangle_{R_3}+\langle TO_3O_3\rangle_{R_4}\right).
	\end{align}
	which gives
	$a_1=c_1+c_2,~a_2=-\frac{c_2}{2}, ~a_3=-\frac{c_3}{2}$.
	
	\subsection{$ \langle J J O_3 \rangle$}
	The WT identity for $\langle JJO_3\rangle$ is given by
	\begin{equation}
		k_{1,\mu}  \langle J^{\mu}(k_1)J^\nu (k_2) O_3(k_3)\rangle =0. 
	\end{equation}
	This implies that only homogeneous solutions exist for conformal ward identity. 
	
	The solution to the homogeneous spinor-helicity ward identities is given by
	\begin{align}
		&\langle J^{-}J^{-}O_3\rangle = \frac{12 (k_1+k_2+2k_3)}{(k_1+k_2+k_3)^2}\langle 12\rangle^2 \quad \langle J^{-}J^{-}O_3\rangle = \frac{12 (k_1+k_2-2k_3)}{(k_1+k_2-k_3)^2}\langle 12\rangle^2\\
		&\langle J^{-}J^{+}O_3\rangle = \frac{12 (-k_1+k_2+2k_3)}{E^2(k_2+k_3-k_1)^2}\langle 31\rangle^2\langle\bar{2}\bar{3}\rangle^2 \quad \langle J^{-}J^{+}O_3\rangle = \frac{12(k_1-k_2+2k_3)}{E^2(k_1-k_2+k_3)^2}\langle 31\rangle^2\langle\bar{2}\bar{3}\rangle^2
	\end{align}
	Notice the homogeneous solutions and their complex conjugates are reproduced in various components of (\ref{basisJJO}). This shows that each of (\ref{basisJJO}) satisfies the homogeneous ward identity.
	This implies that we have four homogeneous solutions which are given by
	\begin{equation}
		\langle JJO_3\rangle_{h_i}=  \langle JJO_3\rangle_{R_i}
	\end{equation}
	with $i=1$ to $4$ and $ \langle JJO_3\rangle_{R_i}$ are given in \eqref{basisJJO}.
	
	The most general solution is then given by 
	\begin{equation}
		\langle JJO_3\rangle= \sum_{i=1}^4 a_i  \langle JJO_3\rangle_{h_i}.
	\end{equation}
	By demanding permutation symmetry between $1$ and $2$ we only get  $a_2=a_3$ that is
	\begin{equation}
		\langle JJO_3\rangle= a_1  \langle JJO_3\rangle_{h_1}+ a_4  \langle JJO_3\rangle_{h_4}+ a_2 \left(  \langle JJO_3\rangle_{h_2}+  \langle JJO_3\rangle_{h_3}\right)
	\end{equation}
	Let us note that, this is more general result than \eqref{fnlansj30}. To obtain this from dS computation one needs to do similar computation as was done in section \ref{mxdgnl1}.

	\subsection{$ \langle TT O_3 \rangle$} 
	The discussion for $\langle TTO_3 \rangle$ is precisely the same as for $\langle JJ O_3 \rangle.$ Again one can show that for $ \langle TT O_3 \rangle$ we  get four homogeneous solutions. By demanding  $1\leftrightarrow 2$ exchange symmetry we obtain
	\begin{equation}\label{tto3gnrl}
		\langle TTO_3\rangle= a_1  \langle TTO_3\rangle_{h_1}+ a_4  \langle TTO_3\rangle_{h_4}+ a_2 \left(  \langle TTO_3\rangle_{h_2}+  \langle TTO_3\rangle_{h_3}\right).
	\end{equation}
	By comparing \eqref{tto3gnrl} and \eqref{tto3gab} we obtain values of $a_1,a_2,a_4$ in terms of $\alpha-$ vacua parameter. Let us note that as compared to results in section \ref{tto3ab} we have more general result. In the special case, results in this section reduces to the results of the section \ref{tto3ab}.

	\subsection{$\langle TTT \rangle$}
	Let us now turn our attention to three point function of stress tensor. 
	The WT identity is given by \eqref{Twtnjh}.
	The three point function can be written as 
	\begin{equation}
		\langle TTT \rangle = \langle TTT \rangle_{h}+ \langle TTT \rangle_{nh}.    
	\end{equation}
	In spinor helcity variables the homogeneous solution for  $\langle TTT\rangle$ is given by
	\begin{align}
		\langle T^{h_1}T^{h_2}T^{h_3}\rangle_{h} = f_{h_{1}, h_{2}, h_{3}}\left(k_{1}, k_{2}, k_{3}\right)\langle 12\rangle^{h_{3}-h_{1}-h_{2}}\langle 23\rangle^{h_{1}-h_{2}-h_{3}}\langle 31\rangle^{h_{2}-h_{3}-h_{1}} 
	\end{align}
	with
	\begin{align}
		&f^{(1)}_{h_{1}, h_{2}, h_{3}}\left(k_{1}, k_{2}, k_{3}\right) =\frac{k_1k_2k_3}{E^{h_1s_1+h_2s_2+h_3s_3}}\bigg|_{s_1 = s_2 = s_3 = 2} \nonumber\\
		&f^{(2)}_{h_{1}, h_{2}, h_{3}}\left(k_{1}, k_{2}, k_{3}\right) \nonumber\\
		&=k_1k_2k_3\left(k_{1}+k_{2}-k_{3}\right)^{h_1s_1+h_2s_2-h_3s_3}\left(k_{2}+k_{3}-k_{1}\right)^{h_3s_3+h_2s_2-h_1s_1}\left(k_{1}+k_{3}-k_{2}\right)^{h_1s_1+h_3s_3-h_2s_2}\bigg|_{s_1 = s_2 = s_3 = 2} \label{homttt}
	\end{align}
	By choosing $h_i=\pm$ one can obtain various helicity components. 
	\subsection{Homogeneous solution: Matching CFT answer with {{${\mathcal W}^3$}} contribution}
	In this subsection we show how to obtain correlation function coming from  {{${\mathcal W}^3$}} term \eqref{hgtttgr}.
	It is interesting to note that, not all the homogeneous solution in \eqref{homttt} is required to reproduce \eqref{hansfull}. The solutions that are required by the $\alpha-$vacua correlator \eqref{hansfull} are as follows
	\begin{align}
		&\langle T^{-}T^{-}T^{-}\rangle_{h,R_1} = f^{(1)}_{---} \langle 12\rangle^2\langle 23\rangle^2\langle 31\rangle^2\notag\\
		&\langle T^{-}T^{-}T^{+}\rangle_{h,R_4} = f^{(2)}_{--+}\frac{\langle 12\rangle^6}{\langle 31\rangle^2\langle 23\rangle^2} \notag\\
		&\langle T^{-}T^{+}T^{-}\rangle_{h,R_3} = f^{(2)}_{-+-}\frac{\langle 31\rangle^6}{\langle 23\rangle^2\langle 12\rangle^2}\notag\\
		&\langle T^{+}T^{-}T^{-}\rangle_{h,R_2} = f^{(2)}_{+--}\frac{\langle 23\rangle^6}{\langle 12\rangle^2\langle 13\rangle^2}\label{homtttsh}
	\end{align}
	By demanding permutation invariance we get \footnote{The BD answer is given by
		$$\langle TTT\rangle_{h,R_1}=\langle TTT\rangle_{h,BD}=F(k_1,k_2,k_3).$$ The alpha vacuum answer should contain all BD vacuum answer as well as other solutions as follows
		\begin{equation}
			F(-k_1,k_2,k_3), F(k_1,-k_2,k_3), F(k_1,k_2,-k_3). 
		\end{equation}
		The permutation invariance then fixes the correlaltor to take the form in \eqref{httt1}.}
	\begin{equation}\label{httt1}
		\langle TTT\rangle_{h} =a  \langle TTT\rangle_{h,R_1}+ b \left(\langle TTT\rangle_{h,R_2}+\langle TTT\rangle_{h,R_3}+\langle TTT\rangle_{h,R_4}\right)
	\end{equation}
	Again we would like to point out, $\alpha-$vacua solution is just a special combination of most general solution that is allowed. In spinor helicity variables, it may look like we have we have total eight independent homogeneous  solutions in \eqref{homttt}, however all of them do not have a good corresponding momentum space representation which is permutation invariant.

	\subsubsection{Non-homogeneous solutions: Matching  contribution due to Einstein term}
	 The non-homogeneous solution to conformal ward identity has to satisfy the ward takahashi identity given in \eqref{Twtnjh}, it can only differ from the BD vacuum by some homogeneous solution. This is simply because the BD vacum answer saturates the WT identity by itself. Since $\alpha-$vacua answer contains the BD answer as well as some additional pieces, these additional pieces must satisfy homogeneous ward identity. 
	 Let us check this expectation explicitly.
	
	In the previous subsection, the homogeneous solutions which  did not play any role in general $\alpha-$vacua  are given by 
	\begin{align}\label{othhno}
		&\langle T^{-}T^{-}T^{-}\rangle_{G_1} = \frac{k_1k_2k_3}{(k_1+k_2-k_3)^2(k_1-k_2+k_3)^2(-k_1+k_2+k_3)^2}\langle 12\rangle^2\langle 23\rangle^2\langle 31\rangle^2 = f^{(2)}_{---}\langle 12\rangle^2\langle 23\rangle^2\langle 31\rangle^2\notag\\
		&\langle T^{-}T^{-}T^{+}\rangle_{G_4} = \frac{k_1k_2k_3}{(k_1+k_2+k_3)^2}\frac{\langle 12\rangle^6}{\langle 23\rangle^2\langle 31\rangle^2}=f^{(1)}_{--+}\frac{\langle 12\rangle^6}{\langle 23\rangle^2\langle 31\rangle^2}\notag\\
		&\langle T^{+}T^{-}T^{-}\rangle_{G_2} = \frac{k_1k_2k_3}{(k_1+k_2+k_3)^2}\frac{\langle 23\rangle^6}{\langle 13\rangle^2\langle 21\rangle^2}=f^{(1)}_{+--} \frac{\langle 23\rangle^6}{\langle 13\rangle^2\langle 21\rangle^2}\notag\\
		&\langle T^{-}T^{+}T^{-}\rangle_{G_3} = \frac{k_1k_2k_3}{(k_1+k_2+k_3)^2}\frac{\langle 13\rangle^6}{\langle 23\rangle^2\langle 21\rangle^2}=f^{(1)}_{-+-}\frac{\langle 13\rangle^6}{\langle 23\rangle^2\langle 21\rangle^2}
	\end{align}
	We now show that these solutions play important role in determining a non-homogeneous solution for the stress tensor three-point function.

	For the general vacuum, we have \eqref{nhabdch}, \eqref{Tnhttg1}
	\begin{align}
		\langle TTT\rangle_{nh, \alpha}-(c_1+c_2) \langle TTT\rangle_{nh, BD} &=-c_2(\langle TTT\rangle_{nh, R_1} - \langle TTT\rangle_{nh, R_2} - \langle TTT\rangle_{nh, R_3} - \langle TTT\rangle_{nh, R_4})
	\end{align} where $\langle TTT\rangle_{nh, R_i}$ appears explicitly in \eqref{nhrit}.
	In the spinor helicity variables we can check it gives
	\begin{align}\label{nhbdaldf}
		&\langle T^{-}T^{-}T^{-}\rangle_{\alpha}-(c_1+c_2) \langle T^{-}T^{-}T^{-}\rangle_{BD} = -96c_2\langle T^{-}T^{-}T^{-}\rangle_{G_1} \notag\\
		&\langle T^{-}T^{-}T^{+}\rangle_{\alpha}-(c_1+c_2) \langle T^{-}T^{-}T^{+}\rangle_{BD} = -96c_2\langle T^{-}T^{-}T^{+}\rangle_{G_4}\notag\\
		&\langle T^{+}T^{-}T^{-}\rangle_{\alpha}-(c_1+c_2) \langle T^{+}T^{-}T^{-}\rangle_{BD} = -96c_2\langle T^{+}T^{-}T^{-}\rangle_{G_2}\notag\\
		&\langle T^{-}T^{+}T^{-}\rangle_{\alpha}-(c_1+c_2) \langle T^{-}T^{+}T^{-}\rangle_{BD} = -96c_2\langle T^{-}T^{+}T^{-}\rangle_{G_3}
	\end{align}
	which are precisely the homogeneous solutions which were left out in  (\ref{homttt}) and are summarised in \eqref{othhno}. We conclude that \eqref{nhbdaldf} precisely matches with our expectation.

	\section{Parity odd contribution to cosmological correlation function in $\alpha-$ vacuum}\label{podd}
	Parity odd contribution to non-gaussianity might also play an important role. In general, one can use the in-in formalism to calculate them just like the parity even case \cite{Maldacena:2011nz, Jain:2021qcl}. For CFT one can also use spin raising and dimension raising operators to calculate them \cite{Jain:2021wyn}. However, very recently it was understood that parity odd CFT three-point function can be obtained by doing epsilon transformation \cite{Caron-Huot:2021kjy, Jain:2021gwa} starting from parity even CFT correlation function. Let us illustrate this in more detail with examples.
	
	\subsection*{$\langle J J O_3 \rangle$}
	To calculate parity odd part of $\langle J J O_3 \rangle$ we need to consider 
	\begin{equation}
		\int \varphi F_{\mu\nu} F_{\rho \sigma} dx^{\mu}\wedge dx^{\nu}\wedge dx^{\rho}\wedge dx^{\sigma}
	\end{equation} in $dS_4$ space. However as was shown in \cite{Jain:2021gwa} the result is given by
	\begin{equation}
		\langle J_{\mu}(k_1) J_{\nu}(k_2) O_3 (k_3) \rangle_{odd}= \frac{1}{k_1}\epsilon_{\mu\alpha\beta}k_{1}^{\beta}\langle J^{\alpha} J_{\nu} O_3 \rangle_{even}.
	\end{equation}
	To obtain the parity odd contribution for $\alpha-$vacua, all we have to do is just plug $\langle J^{\alpha} J_{\nu} O_3 \rangle_{even}$ that appears in \eqref{fnlansj30}.
	
	\subsection*{$\langle T T O_3 \rangle$}For $\langle T T O_3 \rangle$ the story is the same. 
	To calculate parity odd part of $\langle T T O_3 \rangle$ we need to consider 
	\begin{equation}
		\int \varphi {\mathcal W}_{\mu\nu}^{\alpha \beta} {\mathcal W}_{\rho \sigma \alpha \beta} dx^{\mu}\wedge dx^{\nu}\wedge dx^{\rho}\wedge dx^{\sigma}
	\end{equation} in $dS_4$ space. However as was shown in \cite{Jain:2021gwa} the result is given by
	\begin{equation}
		\langle T_{\mu \nu}(k_1) T_{\rho \sigma}(k_2) O_3 (k_3) \rangle_{odd}= \frac{1}{k_1}\epsilon_{( \mu\alpha\beta}k_{1}^{\beta}\langle T^{\alpha}_{ \nu )}(k_1) T_{\rho \sigma}(k_2) O_3 (k_3) \rangle_{even}.
	\end{equation} where we have symmetrised appropriately.
	
	\subsection*{$\langle T T T \rangle$}
	To calculate parity odd part of $\langle T T T \rangle$ we need to consider 
	\begin{equation}
		\int  {\mathcal W}^2 {\tilde W}
	\end{equation}
{ in $dS_4$ space where $W$ is the Weyl tensor, $\tilde{W}$ is the Hodge-dual of the Weyl tensor .} However as was shown in \cite{Jain:2021gwa} the result is given by
	\begin{equation}
		\langle T_{\mu \nu}(k_1) T_{\rho \sigma}(k_2) T_{\gamma \delta} (k_3) \rangle_{odd}= \frac{1}{k_1}\epsilon_{( \mu\alpha\beta}k_{1}^{\beta}\langle T^{\alpha}_{ \nu )}(k_1) T_{\rho \sigma}(k_2)  T_{\gamma \delta} (k_3) \rangle_{even,h}.
	\end{equation}

	\subsection*{General discussion on parity odd contribution in $\alpha-$ vacuum}
	From the CFT perspective, it is very interesting to understand the general structure of parity odd correlation functions in $\alpha-$vacua. In \cite{Giombi:2011rz}, it was shown via some examples that for  correlation function  of conserved currents, when the triangle inequality is violated $s_i+s_j < s_k$ for $i,j,k$ taking value $1,2,3,$ the parity odd contribution is zero. In \cite{Giombi:2016zwa} a more detailed proof of this statement was provided in position space. For example, $\langle T OO \rangle$ can not have parity odd contribution as it violates the triangle inequality. In \cite{Jain:2021whr} a much simpler and intuitive proof was provided and it was shown using momentum space analysis that consistency with OPE  forbids parity odd contribution outside the triangle. However, as explained the $\alpha-$vacua correlators need not be consistent with OPE expansion. This implies one can allow for parity odd contribution even for correlation function outside the triangle such as $\langle T OO \rangle.$ However, these correlators are not present as there are no suitable interactions from the $dS_4$ perspective.

	\section{ $\alpha-$vacua correlator in terms of BD vacum correlator}	\label{albdrl}
	In this section we express results in $\alpha-$vacua,  in terms of BD vacuum answers. 
	Let us start with simplest of cases namely the correlation function in $|\alpha,\beta \rangle$ vacua. For this case the spinning as well as scalar fields the modes in $\alpha-$vacua are defined by the same $\alpha,\beta$ parameters. See section 
	 \ref{thptall} and \ref{spinCFT} for explicit results. Let us start our discussion with the case of scalar three point function. For this case it is easy to see that
	 \begin{align}\label{albd1}
	 \langle O(k_1) O(k_2) O_{k_3}\rangle_{BD}&=F(k_1,k_2,k_3)\nonumber\\
	 \langle O(k_1) O(k_2) O_{k_3}\rangle_{\alpha}&=a~ F(k_1,k_2,k_3) +b\left( F(-k_1,k_2,k_3)+F(k_1,-k_2,k_3)+F(k_1,k_2,-k_3)\right)
	 \end{align}
	 where $a,b$  are given in \eqref{abres}, \eqref{abres1}. For any spinning correlator the relation is the same. The relation can be summarised as follows.
	 Let us define the correlator in BD vacuum to be given by
	\begin{align}\label{albda1}
\sum_{i}  {( \, \rm{distinct~tensor~structure}}\, )_i \, A_i(k_1,k_2,k_3)
	\end{align} 
	 then the answer in $\alpha-$vacua takes the following form
	 \begin{align}\label{albdas1}
	 \sum_{i} {( \, \rm{distinct~tensor~structure}}\, )_i \,  \left[ a ~A_i(k_1,k_2,k_3)+b\left(A_i(-k_1,k_2,k_3)+A_i(k_1,-k_2,k_3)+A_i(k_1,k_2,-k_3)\right)\right]
	 \end{align} 
	 where $A_i(k_1,k_2,k_3)$ are the form factors. This can be checked to be true for all the correlators discussed in \ref{thptall} and \ref{spinCFT} which includes $\langle T OO \rangle, \langle T TO \rangle,\langle T TT \rangle.$
	 
	 Now let us consider more general case when we have $\alpha-$vacua defined as $|\alpha, \beta, {\tilde \alpha},{\tilde \beta} \rangle$, see section \ref{mxdgnl1} for results. For simplicity we focus on $\langle T OO\rangle$ and $\langle TTO\rangle.$ Any other correlators can also be discussed similarly. 
For $\langle TOO\rangle$ we have 
\begin{equation}
\langle TOO\rangle_{BD} = (z_1.k_2)^2 A_T(k_1,k_2,k_3)
\end{equation} 
where the form factor $A_T$ can be found in \eqref{corr1a}. 
For general $\alpha-$vacua we have 
\begin{equation}\label{toognab}
\langle TOO\rangle_{\alpha} = (z_1.k_2)^2 \left( c_1 A_T(k_1,k_2,k_3)+ c_2 A_T(-k_1,k_2,k_3)+c_3\left(A_T(k_1,-k_2,k_3)+A_T(k_1,k_2,-k_3)\right)\right)
\end{equation} 
where $c_i$ are as same as in \eqref{TOOg11}. Let us note that \eqref{toognab} is consistent with $2\leftrightarrow 3$ exchange symmetry. Now let us consider the case of $\langle TTO\rangle.$ For BD vacuum we have
\begin{align}
\langle TTO\rangle_{BD}= \sum_{i} (\,{\rm{distinct~tensor~structure}} \, )_i \, B_i (k_1,k_2,k_3)
\end{align}
where $B_i$ are form factors. The correlator in $\alpha-$vacua in terms of BD vacua is given by
\begin{equation} \label{ttognab}
\begin{split}
\langle TTO\rangle_{\alpha}= \sum_{i} {( \, \rm{distinct~tensor~structure}}\, )_i \,  \bigg[ \, & d_1 ~B_i(k_1,k_2,k_3)+d_2 B_i(k_1,k_2,-k_3) \\ & +d_3\left( B_i(-k_1,k_2,k_3)+B_i(k_1,-k_2,k_3)\right)\bigg]
\end{split}
\end{equation}
which is consistent with \eqref{tto3gab}.
	Let us note that \eqref{ttognab} is consistent with $1\leftrightarrow 2$ exchange symmetry.
	
To conclude, we have shown that given answers in BD vacuum, it is straight forward to obtain correlation function in $\alpha-$vacua.

	\section{Discussion}\label{dis}
	In this paper, we have discussed the cosmological correlation function in general $\alpha$-vacua in rigid dS space. One of the main purposes of this paper is to understand how to construct these cosmological correlation functions from a CFT perspective. We showed that for this purpose we need to consider CFT correlators that are not consistent with OPE limit. Interestingly, conformal ward identity in momentum space allows for such solutions. For example, solving conformal ward identity for the three-point function of scalar operator gives in general four different solutions. Out of these four different solutions, only one of them is consistent with OPE and coincides with the correlation function in the Bunch-Davies vacuum as well as consistent with the position space correlation function. However, for $\alpha-$vacua all the four solutions are important and the most general solution is a linear combination of all these four solutions consistent with permutation symmetry. For the spinning correlator, we then used spin and dimension raising operators as well as a solution of conformal ward identity in momentum/spinor helicity variables.  Based on our computation we summarise the result as follows
	\begin{itemize}
		\item BD vacuum answers $\implies$ Imposing consistency with OPE limit and permutation invariance on solution to conformal ward identity in momentum space.
		\item $\alpha-$vacua answer ~$\implies$ Relaxing consistency with OPE limit however  keeping consistency with permutation invariance  on solution to conformal ward identity in momentum space.
	\end{itemize}
	
	There are several important aspects of these analyses which we plan to explore in future.
	\subsection*{Inflationary correlation function}
	In this paper we have focused on calculation of three point function in rigid dS space. It would be interesting to calculate the correlation function for inflationary case. For rigid de-Sitter case we have shown how to obtain correlation function in $\alpha-$vacua given correlation function BD vaccum, see \ref{albdrl}.
	It will be interesting to check if the same relation continues to hold in the inflationary scenario. This will also provide an easy way to obtain answers in $\alpha-$vacua given plethora of results  already known in the BD vacuum.
	\subsection*{Observational significance}  
	In this paper we have mostly concentrated on formal aspects of correlation functions. We have  not studied any phenomenological implication of the $\alpha-$vacua. It is well known that signal for the $\alpha-$vacua can be significantly enhanced as compared to BD vacuum, see \cite{Kanno:2022mkx} for a recent discussion on this issue. We have also not studied the issue of consistency of  $\alpha-$vacua. One of the problems with $\alpha-$vacua is that  stress tensor expectation value of probe scalar field is divergent and as a result, its back-reaction can invalidate the rigid dS approximation.  To renormalize one needs non-local and $\alpha-$dependent counter-term. This seems to be problematic, however,
	it was argued in \cite{Goldstein:2003ut} that $\alpha-$parameter-dependent counter-terms are fine. We would like to come back to this issue in the future.
	
	In general, it would be nice to understand CFT's by relaxing OPE consistency conditions and if one can realise these kind of CFT's in some other interesting  physical system.
	
	\section*{Acknowledgements} The work of S.J  is supported by the Ramanujan Fellowship. AM would like to acknowledge the support of CSIR-UGC
	(JRF) fellowship (09/936(0212)/2019-EMR-I). We would like to thank S. P. Trivedi for collaboration and extensive discussion on the content of the draft. We thank S. Minwalla, D. Ghosh, and S. Mukhi for their helpful discussions. SJ would like to thank DTP, TIFR for providing excellent hospitality during the course of the work.
	We acknowledge our debt to the people of India for their steady support of research in basic sciences.

	\appendix
	\section*{Appendix}
	\section{Consistency with OPE limit}
	In this section we follow discussion in section $2.1$ of \cite{Jain:2021whr}.
	In momentum space, there are four solutions to conformal ward identity for scalar three point function. One can show that all these four solutions can be combined to give most general correlation function to be given by
	\begin{align} 
		\langle O O O\rangle=a_{1} f\left(k_{1}+k_{2}+k_{3}\right)+a_{2} f\left(-k_{1}+k_{2}+k_{3}\right)+a_{3} f\left(k_{1}-k_{2}+k_{3}\right)+a_{4} f\left(k_{1}+k_{2}-k_{3}\right).
	\end{align}
	Permutation symmetry implies that $a_2 = a_3 = a_4$.
	Let us consider simple case of scalar operator $O$ with $\Delta = 2.$ For this case 
	we have $f(k_1+k_2+k_3) = \ln(k_1+k_2+k_3)$. 
	The same correlator in position space in the 
	OPE limit, $x_{23}\to 0$,  goes like  $\langle OOO\rangle \sim \frac{1}{x^2_{23}x^4_{12}}$ which in momentum space leads to $\langle OOO\rangle \sim \frac{k_1}{k_3}$ with $k_2 \approx k_3 \gg k_1$ where $k_1 \to 0.$ It is easy to check that this can be only be reproduced by $\ln(k_1+k_2+k_3)$. 
	Hence, singularity of the form  $f(k_i-k_j+k_k)$ is not consistent with the OPE limit and only singularity structure $E = k_1 + k_2 + k_3 \rightarrow 0$ is consistent. 
	So we conclude that consistency with OPE limit restricts the correlator to only have a total energy pole that is a pole in $E=k_1+k_2+k_3.$

	\section{Shadow Transform in dS correlator}\label{Shad}
	In dS, correlators for boundary operators are related to the correlators of the bulk field through shadow transform. Though this is a standard and well-known technique\cite{Maldacena:2011nz, Mata:2012bx}, we will elaborate with a simple example for completeness.\\
	
	Consider a bulk field $\phi(x)$ in dS background and $O(x)$ is the corresponding dual operator in the `boundary CFT'. The wave-function of the universe (using in-in formalism) is written in terms of this bulk field as
	\begin{equation}\label{WFUinin}
		\begin{split}
			\Psi[\phi(x)] = \exp\left( -\frac{1}{2!} \int d^3x d^3y \phi(x)\phi(y) \langle O(x)O(y) \rangle +\frac{1}{3!} \int d^3x d^3y d^3z \phi(x)\phi(y)\phi(z) \langle O(x)O(y)O(z) \rangle +\cdots\right)
		\end{split}
	\end{equation}
	
	The interpretation of this wave function is that one can access all information about the dynamics of this bulk field which can be related to boundary operator $O(x)$ with dimension $\Delta$. This similar fact is also seen in more conventional shadow transform in AdS correlators. One can calculate various moments of the bulk field using,
	

	\begin{equation}\label{Wvfn}
		\begin{split}
			\langle \phi_{\vec{k}_1}\cdots\phi_{\vec{k}_n}\rangle = \frac{\int \mathcal{D}\phi~ \phi_{\vec{k}_1}\cdots\phi_{\vec{k}_n} |\Psi[\phi]|^2}{\int \mathcal{D}\phi |\Psi[\phi]|^2}
		\end{split}
	\end{equation}
	which reduces to the in-in formalism correlator (eq \eqref{ininf}). Using perturbation of the bulk field we can obtain series of relations between moments of $\phi(x)$ and the set of correlators of the boundary dual insertion $O(x)$. Below we will state the relations
	
	\begin{equation}
		\begin{split}
			\langle\phi_{\vec{k}_1}\phi_{\vec{k}_2}\rangle &= \frac{1}{2 {\rm Re}\langle O_{\vec{k}_1} O_{\vec{k}_2} \rangle}\\
			\langle\phi_{\vec{k}_1}\phi_{\vec{k}_2}\phi_{\vec{k}_3}\rangle &= \frac{{\rm Re}\langle O_{\vec{k}_1} O_{\vec{k}_2} O_{\vec{k}_3} \rangle}{2 \prod_{i=1}^3{\rm Re}\langle O_{\vec{k}_i} O_{-\vec{k}_i} \rangle}\\
			\cdots
		\end{split}
	\end{equation}
	
	The two function above is obtained by doing a standard Gaussian path integral over the bulk field. On the other hand, the other side is obtained using the expansion of the wave function up to cubic order and using the standard rules of wick contraction.\\
	
	Similar kinds of relations can be obtained for higher spin field correlators. Also, the above relations can be inverted systematically to write boundary correlators in terms of bulk field correlators.\\

	\section{Some back ground details}\label{bckd}

	In this section, we collect some background details which are useful for the paper.
	\subsection{Momentum space representation}
	To compute a correlation function in momentum space CFT, one needs to use the momentum space representation of the generators where the ward identities are differential equations given by\cite{Bzowski:2013sza}
	\begin{align}
		0=\left[\sum_{j=1}^{n} \Delta_{j}-(n-1) d-\sum_{j=1}^{n-1} p_{j}^{\alpha} \frac{\partial}{\partial p_{j}^{\alpha}}\right]\left\langle\left\langle\mathcal{T}_{1}\left(\boldsymbol{p}_{1}\right) \ldots \mathcal{T}_{n}\left(\boldsymbol{p}_{n}\right)\right\rangle\right\rangle
	\end{align}
	
	\begin{align}
		\left.0=\left[\sum_{j=1}^{n-1}\left(2\left(\Delta_{j}-d\right) \frac{\partial}{\partial p_{j}^{\kappa}}-2 p_{j}^{\alpha} \frac{\partial}{\partial p_{j}^{\alpha}} \frac{\partial}{\partial p_{j}^{\kappa}}+\left(p_{j}\right)_{\kappa} \frac{\partial}{\partial p_{j}^{\alpha}} \frac{\partial}{\partial p_{j \alpha}}\right)\right]\left\langle\mathcal{O}_{1}\left(\boldsymbol{p}_{1}\right) \ldots \mathcal{O}_{n}\left(\boldsymbol{p}_{n}\right)\right\rangle\right\rangle
	\end{align}
	with additional contribution for spinning operators
	\begin{align}
		\begin{aligned}
			&2 \sum_{j=1}^{n-1} \sum_{k=1}^{n_{j}}\left(\delta^{\mu_{j k} \kappa} \frac{\partial}{\partial p_{j}^{\alpha_{j k}}}-\delta_{\alpha_{j k}}^{\kappa} \frac{\partial}{\partial p_{j \mu_{j k}}}\right) \times \\
			&\quad \times\left\langle\left\langle\mathcal{T}_{1}^{\mu_{11} \ldots \mu_{1 r_{1}}}\left(\boldsymbol{p}_{1}\right) \ldots \mathcal{T}_{j}^{\mu_{j 1} \ldots \alpha_{j k} \ldots \mu_{j r_{j}}}\left(\boldsymbol{p}_{j}\right) \ldots \mathcal{T}_{n}^{\mu_{n 1} \ldots \mu_{n r_{n}}}\left(\boldsymbol{p}_{n}\right),\right\rangle\right\rangle
		\end{aligned}
	\end{align}
	Since the two-point functions and three-point functions are completely determined in a  CFT by the Ward identities, for $n \leq 3$ these differential equations can be solved exactly\footnote{And uniquely, provided OPE consistency is implemented.} up to an overall constant.

	\subsection{Spinor-helicity representation}\label{sphrep11}
	It is sometimes usefult to consider the spinor-helicity representation for the CFT generators. A $3D$ momentum variable may be decomposed as into $SL(2, \mathbb{C})$ via\cite{Baumann:2020dch}
	\begin{align}
		\lambda_{\alpha}\bar{\lambda}^{\beta} = k_i(\sigma^i)_{\alpha}^{~\beta}+ k \mathbb{I}_{\alpha}^{~\beta}
	\end{align}
	In the $SL(2, \mathbb{C})$ variables the ward identity looks like
	\begin{align}
		\sum^n_{a=1}\widetilde{K}_a^{\kappa}\left\langle\frac{J_{s_{1}}}{k_{1}^{s_{1}-1}} \frac{J_{s_{2}}}{k_{2}^{s_{2}-1}} \cdots \frac{J_{s_{n}}}{k_{n}^{s_{n}-1}}\right\rangle=\text { transverse Ward identity terms }
	\end{align}
	where
	\begin{align}
		\tilde{K}^{i}=2\left(\sigma^{i}\right)_{\alpha}^{\beta} \frac{\partial^{2}}{\partial \lambda_{\alpha} \partial \bar{\lambda}^{\beta}}
	\end{align}
	is the generator of special conformal transfomations in the spinor-helicity representation. The above ward identities prompt the following decomposition of all correlators\cite{Jain:2021vrv}
	\begin{align}
		\left\langle J_{s_{1}}J_{s_{2}}\cdots J_{s_{n}}\right\rangle = \left\langle J_{s_{1}}J_{s_{2}}\cdots J_{s_{n}}\right\rangle_{h}+\left\langle J_{s_{1}}J_{s_{2}}\cdots J_{s_{n}}\right\rangle_{nh}
	\end{align}
	where
	\begin{align}
		&\sum^n_{a=1}\widetilde{K}_a^{\kappa}\left\langle\frac{J_{s_{1}}}{k_{1}^{s_{1}-1}} \frac{J_{s_{2}}}{k_{2}^{s_{2}-1}} \cdots \frac{J_{s_{n}}}{k_{n}^{s_{n}-1}}\right\rangle_h = 0\\
		&\sum^n_{a=1}\widetilde{K}_a^{\kappa}\left\langle\frac{J_{s_{1}}}{k_{1}^{s_{1}-1}} \frac{J_{s_{2}}}{k_{2}^{s_{2}-1}} \cdots \frac{J_{s_{n}}}{k_{n}^{s_{n}-1}}\right\rangle_{nh}=\text { transverse Ward identity terms }
	\end{align}
	\subsection{Spin-raising and Weight-shifting operators}\label{spinrai}
	In momentum space, it is possible to define Weight-shifting and Spin-raising operators that raise or lower the spin or weight of external operators in a correlation function without violating the CFT Ward identities. Such operators are very useful in generating higher spin correlations from a scalar seed. All spin-raising operators are schematically\cite{Baumann:2019oyu}
	\begin{align}
		\mathcal{O}: F_{\{\Delta_i, J_i\}} \to F_{\{\Delta'_i, J'_i\}}
	\end{align}
	where $F_{\{\Delta_i, J_i\}}$ are correlation functions satisfying 
	\begin{align}
		\delta_{\omega}F_{\{\Delta_i, J_i\}} = 0
	\end{align}
	Below we state some of the simple spin-raising and weight-shifting operators
	\begin{align}
		&\mathcal{W}^{--}_{12}: F_{\Delta_1, \Delta_2} \to F_{\Delta_1-1, \Delta_2-1}\notag\\
		&\mathcal{W}^{--}_{12} = \frac{1}{2}\left(\frac{\partial}{\partial \vec{k}_1}-\frac{\partial}{\partial \vec{k}_2}\right)\cdot \left(\frac{\partial}{\partial \vec{k}_1}-\frac{\partial}{\partial \vec{k}_2}\right) \equiv \frac{1}{2}\vec{K}_{12}\cdot \vec{K}_{12}\\\notag\\
		&\mathcal{W}^{++}_{12}: F_{\Delta_1, \Delta_2} \to F_{\Delta_1+1, \Delta_2+1}\notag\\
		&\mathcal{W}_{12}^{++}=\left(k_{1} k_{2}\right)^{2} \mathcal{W}_{12}^{--}-\left(d-2 \Delta_{1}\right)\left(d-2 \Delta_{2}\right) \vec{k}_{1} \cdot \vec{k}_{2} \notag\\
		&+\left(k_{2}^{2}\left(d-2 \Delta_{1}\right)\left(d-1-\Delta_{1}+\vec{k}_{1} \cdot \vec{K}_{12}\right)+(1 \leftrightarrow 2)\right)\\\notag\\
		&\mathcal{S}^{++}_{12}: F_{J_1, J_2} \to F_{J_1+1, J_2+1}\notag\\
		&\mathcal{S}^{++}_{12}=\left(S_{1}+\Delta_{1}-1\right)\left(S_{2}+\Delta_{2}-1\right) \vec{z}_{1} \cdot \vec{z}_{2}-\left(\vec{z}_{1} \cdot \vec{k}_{1}\right)\left(\vec{z}_{2} \cdot \vec{k}_{2}\right) \mathcal{W}_{12}^{--} \notag\\
		&+\left[\left(S_{1}+\Delta_{1}-1\right)\left(\vec{k}_{2} \cdot \vec{z}_{2}\right)\left(\vec{z}_{1} \cdot \vec{K}_{12}\right)+(1 \leftrightarrow 2)\right]\\\notag\\
		&D_{12}: F_{\Delta_1, J_1, \Delta_2, J_2} \to F_{\Delta_1, J_1+1, \Delta_2-1, J_2}\notag\\
		&D_{12} = (\Delta_1+S_1-1)\vec{z}_1\cdot \vec{K}_{12} - (\vec{z}_1\cdot \vec{k}_1)\mathcal{W}^{--}_{12}\\\notag\\
		&H_{12}: F_{\Delta_1, J_1, \Delta_2, J_2} \to F_{\Delta_1-1, J_1+1, \Delta_2-1, J_2+1}\notag\\
		&H_{12} = 2\vec{z}_2\cdot \vec{K}_{12}\vec{z}_2\cdot \vec{K}_{12} - 2(\vec{z}_1\cdot \vec{z}_2)\mathcal{W}^{--}_{12}
	\end{align}

	\section{Weyl tensor}\label{weylt}
	For the $\alpha-$vacua, the $dS$ Weyl tensor is related to the flat space via
	\begin{align}
		[{\mathcal W}^{dS}]^{\mu}_{\nu\rho\sigma} = -Cik\eta [{\mathcal W}^{flat}]^{\mu}_{\nu\rho\sigma}(e^{ik\eta+i\vec{k}.\vec{x}})+Dik\eta [{\mathcal W}^{flat}]^{\mu}_{\nu\rho\sigma}(e^{-ik\eta+i\vec{k}.\vec{x}})
	\end{align}
	To show the above, consider the Weyl tensor at linearized order,
	\begin{align}
		{\mathcal W}^{\mu}_{\nu\alpha\beta} = \frac{1}{2}[\gamma_{\mu\alpha,\nu\beta}-\gamma_{\mu\beta,\nu\alpha}-\gamma_{\nu\alpha,\mu\beta}+\gamma_{\nu\beta, \mu\alpha}]
	\end{align}
	Now, it can be shown that
	\begin{align}
		\dot{\gamma}_{ab} = \frac{\partial}{\partial\eta}\gamma_{ab}(k, \eta) = \frac{z_az_b}{\sqrt{2k^3}}k^2\eta[e^{ik\eta}C -  e^{-ik\eta}D]\label{td}
	\end{align}
	Now, since, the non-zero components of the linearized Weyl tensor is
	\begin{align}
		{\mathcal W}_{0 i0 j} &=\frac{1}{2} \ddot{\gamma}_{i j} \notag\\
		{\mathcal W}_{i j0 k} &=\frac{1}{2}\left(\dot{\gamma}_{k i, j}-\dot{\gamma}_{k j, i}\right) \notag\\
		{\mathcal W}_{0 i j k} &=\frac{1}{2}\left(\dot{\gamma}_{i k, j}-\dot{\gamma}_{i j, k}\right) \notag\\
		{\mathcal W}_{i j k l} &=\frac{1}{2}\left(\gamma_{i k, jl} -\gamma_{i l, jk}-\gamma_{j k, il} +\gamma_{j l, i k}\right)
	\end{align}
	Using (\ref{td}) in the above, we can immediately see that
	\begin{align}
		[{\mathcal W}^{dS}]^{\mu}_{\nu\rho\sigma} = -Cik\eta [{\mathcal W}^{flat}]^{\mu}_{\nu\rho\sigma}(e^{ik\eta+i\vec{k}.\vec{x}})+Dik\eta [{\mathcal W}^{flat}]^{\mu}_{\nu\rho\sigma}(e^{-ik\eta+i\vec{k}.\vec{x}}).\label{meW}
	\end{align}
	This means that the Weyl tensor obeys the mode expansion
	\begin{align}
		[{\mathcal W}^{dS}]^{\mu}_{\nu\rho\sigma}(k, \eta) = \frac{1}{\sqrt{2k^3}}[-Cik\eta e^{ik\eta}[{\mathcal W}^{+}]^{\mu}_{\nu\rho\sigma}+Dik\eta e^{-ik\eta}[{\mathcal W}^{-}]^{\mu}_{\nu\rho\sigma}]
	\end{align}
	where $\pm$ denotes the use of $(\pm k, k^i)$ in the flat-space Weyl tensors, respectively. 
	We also define
	\begin{align}
		&v^{\alpha\beta\gamma\delta}_{k}(\eta) = \frac{1}{\sqrt{k^3}}[e^{ik\eta}(-i k\eta)[{\mathcal W}^{+}]_{\alpha\beta\gamma\delta}A - B e^{-ik\eta}(i k\eta)[{\mathcal W}^{-}]_{\alpha\beta\gamma\delta}]\label{meW2}\\
		& [{\mathcal W}^{+}]_{\alpha\beta\gamma\delta}^{\dagger} = [{\mathcal W}^{-}]_{\alpha\beta\gamma\delta}
	\end{align}
	where 
	\begin{align}
		{\mathcal W}^{\pm}_{\eta\zeta\alpha\beta}(k^{\mu}) \equiv {\mathcal W}_{\eta\zeta\alpha\beta}(\pm k, \vec{k}).
	\end{align}

	\section{Cosmological correlation function}
	In this Appendix, we compute several cosmological correlation functions which play important role in the main text. 
	\subsection{$\langle TOO\rangle$}\label{gtoob}
	To compute two scalars and one graviton amplitude we need to consider 
	\begin{align}
		H_{int} = \int d^4x \sqrt{-g}~g^{\mu\nu}\partial_{\mu}\phi\partial_{\nu}\phi.    \end{align}
	Using the above general vacua mode expansion \eqref{me},\eqref{modegravi1},  along with  (\ref{ininf}) we obtain 
	the following time integral 
	\begin{align}
		(z_1.k_2)^2\Im\bigg[&\int_{-\infty}^0\frac{d\eta}{\eta^2} \bigg(-2f_{k_3}(A, B)u_{k_3}(\eta)[\bar{f}_{k_1}(C, D)f_{k_2}(A, B) \bar{\gamma}_{k_1}(\eta)\bar{u}_{k_2}(\eta)\nonumber\\
		&-2f_{k_1}(C,D)f_{k_2}(A, B)\gamma_{k_1}(\eta)u_{k_2}(\eta)]\nonumber\\
		&+2\bar{f}_{k_1}(C, D)\bar{f}_{k_2}(A, B)\bar{\gamma}_{k_1}(\eta)\bar{u}_{k_2}(\eta)[-\bar{f}_{k_3}(A, B) \bar{u}_{k_3}(\eta)+f_{k_3}(A, B) u_{k_3}(\eta)]\bigg)\bigg]
	\end{align}
	where $f_k(\alpha)$ and $\gamma_k(\eta)$ has been defined in (\ref{varfun}) and (\ref{gamma}) previously. The $\langle TOO\rangle$ correlator can be found after computing the time integral and performing the shadow transform can be shown to be given by
	\begin{align}
		\langle TO_3O_3\rangle = c_1\langle TO_3O_3\rangle_1 +c_2\langle TO_3O_3\rangle_2+c_3 \left(\langle TO_3O_3\rangle_3+\langle TO_3O_3\rangle_4\right)\label{TOOg}
	\end{align}
	where $c_i$ are given by
	\begin{align}
		&c_1 = \Re\left\{\frac{1}{2(C+D)(\bar{A}+\bar{B})^2}\left[\frac{\bar{A}^2+\bar{B}^2}{\bar{C}+\bar{D}}+\frac{((2B D+A(C+D))\bar{A}+(2AC+B(C+D))\bar{B})}{(A+B)^2}\right]\right\}\nonumber\\
		&c_2 = \Re\left\{\frac{1}{2(C+D)(\bar{A}+\bar{B})^2}\left[-\frac{\bar{A}^2+\bar{B}^2}{\bar{C}+\bar{D}}+\frac{(2B C+A(C+D))\bar{A}+(2AD+B(C+D))\bar{B}}{(A+B)^2}\right]\right\}\nonumber\\
		&c_3 = \Re\left\{\frac{1}{2(C+D)(\bar{A}+\bar{B})^2}\left[\frac{2\bar{A}\bar{B}}{\bar{C}+\bar{D}}-\frac{(-B\bar{A}+A\bar{B})(C-D)}{(A+B)^2}\right]\right\}\label{abc}
	\end{align}
	For the special case when we consider $A=C, B=D$ we have
	\begin{align}
		\langle TO_3O_3\rangle = a\langle TO_3O_3\rangle_1 +b\left(\langle TO_3O_3\rangle_2+\langle TO_3O_3\rangle_3+\langle TO_3O_3\rangle_4\right)\label{TOOgg1}  
	\end{align}
	where the form of $a,b$ in \eqref{TOOgg1} are given by
	\begin{align}\label{solabtoo}
		a&=\frac{1}{2\mathcal{N}^3(A, B)}[(2A^2+3 AB+3B^2)(\bar{A}^2+\bar{B}^2+\bar{A}\bar{B})+(A-B)\bar{B}(|A|^2+|B|^2+A\bar{B})]\nonumber\\
		b&=\frac{1}{2\mathcal{N}^3(A, B)}[A(A-B)B(\bar{A}^3+\bar{B}^3)+(A^3+6A^2B+ B^3)(\bar{A}^2\bar{B}-\bar{A}\bar{B}^2)]
	\end{align}
	and 
	\small
	\begin{align}
		&\langle TO_3O_3\rangle_{R_1} =\left[k_1+k_2+k_3-\frac{k_1k_2+k_2k_3+k_3k_1}{k_1+k_2+k_3}-\frac{k_1k_2k_3}{(k_1+k_2+k_3)^2}\right] (z_1.k_2)^2\nonumber\\
		&\langle TO_3O_3\rangle_{R_2} = \left[-k_1+k_2+k_3-\frac{-k_1k_2+k_2k_3-k_3k_1}{-k_1+k_2+k_3}+\frac{k_1k_2k_3}{(-k_1+k_2+k_3)^2}\right](z_1.k_2)^2\nonumber\\
		&\langle TO_3O_3\rangle_{R_3} =  \left[k_1-k_2+k_3-\frac{-k_1k_2-k_2k_3+k_3k_1}{k_1-k_2+k_3}+\frac{k_1k_2k_3}{(k_1-k_2+k_3)^2}\right](z_1.k_2)^2\nonumber\\
		&\langle TO_3O_3\rangle_{R_4} = \left[k_1+k_2-k_3-\frac{k_1k_2-k_2k_3-k_3k_1}{k_1+k_2-k_3}+\frac{k_1k_2k_3}{(k_1+k_2-k_3)^2}\right](z_1.k_2)^2\label{corr1a11}
	\end{align}
	
	The Ward identity for a correlator in the above form is by brute computation can be shown to be
	\begin{align}\label{WTob}
		k^{\mu}_1z^{\nu}\langle T_{\mu\nu} O_3 O_3\rangle &=(a+b)(z_1.k_2) (\langle O_3(k_2)O_3(-k_2)\rangle_{BD}-\langle O_3(k_3)O_3(-k_3)\rangle_{BD}) \nonumber\\
		&= \frac{1}{(A+B)(\bar{A}+\bar{B})} (z_1.k_2) (\langle O_3(k_2)O_3(-k_2)\rangle_{BD}-\langle O_3(k_3)O_3(-k_3)\rangle_{BD}) 
	\end{align}
	where we have used \eqref{solabtoo} to obtain
	\begin{align}
		a+b = \frac{1}{(A+B)(\bar{A}+\bar{B})}
	\end{align}
	Let us note that WT identity should be consistent with
	\begin{align}\label{WTd}
		k^{\mu}_1z^{\nu}\langle T_{\mu\nu} O_3 O_3\rangle &=(z_1.k_2)  (\langle O_3(k_2)O_3(-k_2)\rangle_{\alpha}-\langle O_3(k_3)O_3(-k_3)\rangle_{\alpha}) \nonumber\\
		&= \frac{1}{(A+B)(\bar{A}+\bar{B})} (z_1.k_2) (\langle O_3(k_2)O_3(-k_2)\rangle_{BD}-\langle O_3(k_3)O_3(-k_3)\rangle_{BD}) 
	\end{align}
	where in the last line we have used (\ref{2pt}) i.e.
	\begin{align}\label{2pt1}
		\langle O(k_2)O(-k_2)\rangle_{\alpha} = \frac{1}{(A+B)(\bar{A}+\bar{B})} \langle O(k_2)O(-k_2)\rangle_{BD}
	\end{align}
	We observe that  \eqref{WTd} with  \eqref{WTob} are identical as should be the case. Alternative one can use this consistency condition to fix the normalization of $A,B$ to be as given in 
	\eqref{conAB}
	\footnote{Note that the CFT ward identity constraints the $A, B$. 
		This makes the connection to alpha vacuum clear from CFT side. 
		In position space, $\langle TOO\rangle$ is given by (see \cite{Osborn:1993cr})
		\begin{align}
			\langle T_{\mu\nu}(x_2)O_{\Delta}(x_2)O_{\Delta}(x_3)\rangle = \frac{1}{x^d_{12}x^{2\Delta-d}_{23}x^d_{31}}\frac{d N\Delta}{(1-d)S_d} h^1_{\mu\nu}(\hat{X}_{23})
		\end{align}
		where
		\begin{align}
			h^1_{\mu\nu} = \hat{X}_{23 \mu}\hat{X}_{23 \nu}-\frac{1}{d}\delta_{\mu\nu} \quad \hat{X}_{23\mu} = \frac{X_{23\mu}}{X^2_{23}}, X_{23} = \frac{x_{21}}{x^2_{21}}-\frac{x_{31}}{x^2_{31}}
		\end{align}
		and 
		\begin{align}
			\langle O_{\Delta}(x_1)O_{\Delta}(x_2)\rangle = \frac{N}{x^{2\Delta}_{12}}
		\end{align}
		So the three-point function OPE coefficient is determined in terms of the two-point function $N.$ This is a reflection of the fact that $\langle TOO\rangle $ is non-homogeneous.}.

	\subsection{ $\langle TTO_3\rangle$}\label{TTobg}
	To calculate two graviton and one scalar amplitude
	we consider the following interaction term
	\begin{equation}\label{inttto1}
		H_{int}=\int d^4x~\sqrt{-g}\varphi {\mathcal W}_{\rho\sigma\alpha \beta} {\mathcal W}_{\rho \sigma \alpha \beta}
	\end{equation} 
	Using the mode expansion \eqref{modegravi1} along with \eqref{me} in \eqref{inttto1}
	the following time integral can be obtained
	\begin{align}
		\Im\bigg[&\int_{-\infty}^0d\eta [2\bar{f}_{k_1}(C, D)v^{\alpha\beta\mu\nu}_{k_1}(\eta)(\bar{f}_{k_2}(C, D)v^{\alpha\beta\mu\nu}_{k_2}(\eta)-f_{k_2}(C, D)\bar{v}^{\alpha\beta\mu\nu}_{k_2}(\eta))\notag\\&+(\bar{f}_{k_2}(C, D)v^{\alpha\beta\mu\nu}_{k_2}(\eta)+f_{k_2}(C, D)\bar{v}^{\alpha\beta\mu\nu}_{k_2}(\eta))(\bar{f}_{k_1}(C, D)v^{\alpha\beta\mu\nu}_{k_1}(\eta)-f_{k_1}(C, D)\bar{v}^{\alpha\beta\mu\nu}_{k_1}(\eta))]\notag\\
		&(\bar{f}_{k_3}(A, B)u_{k_3}(\eta)-f_{k_3}(A, B)\bar{u}_{k_3}(\eta))\bigg]
	\end{align}
	where $f_{k}$ and $u_{k}(\eta)$ has been defined in (\ref{varfun})and $v^{\alpha\beta\mu\nu}_k(\eta)$ has been defined in (\ref{meW2}). See appendix \ref{weylt} for details on Weyl tensor in terms of mode expansion. The above gives
	\begin{align}\label{tto3gab11}
		\langle TTO_3\rangle_{\alpha} = d_1\langle TTO_3\rangle_{R_1}+d_2\langle TTO_3\rangle_{R_4}+ d_3(\langle TTO_3\rangle_{R_2} + \langle TTO_3\rangle_{R_3})
	\end{align}
	where
	\begin{align}\label{di123}
		&d_1 = \Re\left[\frac{-|B|^2C^2(\bar{C}+\bar{D})^2+\bar{A}-B|C|^4-2BC\bar{D}|C|^2+(A(C+D)^2+BD(2C+D)\bar{D}^2)]}{6(A+B)(C+D)^2(\bar{A}+\bar{B})(\bar{C}+\bar{D})^2}\right]\nonumber\\
		&d_2 = a(A\leftrightarrow B)\nonumber\\
		&d_3 =\Re\left[\frac{\left((A-B)CD(\bar{A}+\bar{B})(\bar{C}^2+\bar{D}^2)+ \textbf{c.c.}\right)+4|C|^2 |D|^2(|A|^2-|B|^2)}{6(A+B)(C+D)^2(\bar{A}+\bar{B})(\bar{C}+\bar{D})^2}\right]
	\end{align}
	and $\langle TTO_3\rangle_{R_i}$ has been defined in (\ref{avac2}) and (\ref{avac3}). Notice, if we put $C = A, D = B$, then, we find
	\begin{align}
		d_1=a \quad  d_2 = d_3=b
	\end{align}
	where $a,b$ are same as that appeard in \eqref{cftans11}.
	To summarise for this special case we have 
	\begin{align}
		\langle TTO_3\rangle_{\alpha} = c_1\langle TTO_3\rangle_{R_1}+c_2\left(\langle TTO_3\rangle_{R_4}+\langle TTO_3\rangle_{R_2} + \langle TTO_3\rangle_{R_3}\right).
	\end{align}
	The explicit expressions for $\langle TTO_3 \rangle_{R_i}$ are complicated. They are best written in spinor helicity variables. We give their explicit forms in section \ref{tto3ab}.

	\subsection{$\langle TTT\rangle$ }\label{TTTgb1}
	The three graviton amplitude can get contribution from two different sources, the Einstein Hilbert term, and the $Weyl^3$  term. Let us start with the Weyl tensor contribution.
	\subsubsection*{$Weyl^3$ contribution}
	To calculate the three graviton amplitude we need to consider the following interaction 
	\begin{align}
		S^{(3)}_{\gamma, {\mathcal W}^3} = \int d^4x\sqrt{-g}~ {\mathcal W}^3 \label{hI}.
	\end{align}
	Using the mode expansion \eqref{modegravi1}  in \eqref{inttto1}
	the following time integral can be obtained
	\begin{align}
		& \Im\bigg[\int d\eta ~\eta^2~ (-f_{k_3}(C, D) v^{\alpha\beta\gamma\delta}_{k_3}(\eta)+\bar{f}_{k_3}(C, D) v^{\alpha\beta\gamma\delta}_{k_3}(\eta) f_{k_2}(C, D)f_{k_1}(C, D) \bar{v}^{\gamma\delta\eta\zeta}_{k_2}(\eta)\bar{v}^{\eta\zeta\alpha\beta}_{k_1}(\eta)\notag\\&+(-f_{k_2}(C, D) \bar{v}^{\alpha\beta\gamma\delta}_{k_2}(\eta)+\bar{f}_{k_2}(C, D)v^{\alpha\beta\gamma\delta}_{k_2})(\eta)|f(C, D)|^2\bar{v}^{\gamma\delta\eta\zeta}_{k_1}(\eta)v^{\eta\zeta\alpha\beta}_{k_3}(\eta)\notag\\&+(-f_{k_1}(C, D) v^{\alpha\beta\gamma\delta}_{k_1}(\eta)+\bar{f}_{k_1}(C, D) v^{\alpha\beta\gamma\delta}_{k_1}(\eta))\bar{f}_{k_2}(C, D)\bar{f}_{k_3}(C, D) v^{\gamma\delta\eta\zeta}_{k_2}(\eta)v^{\eta\zeta\alpha\beta}_{k_3}(\eta)\bigg]\label{Tittt}
	\end{align}
	where $f_k(\alpha)$ has been defined in (\ref{varfun}) and  $v^{\alpha\beta\gamma\delta}_{k}(\eta)$ is defined in \eqref{meW2}. See appendix \ref{weylt} for details on Weyl tensor in terms of mode expansion.
	We also have 
	\begin{align}
		&v^{\alpha\beta\gamma\delta}_{k}(\eta) = \frac{1}{\sqrt{k^3}}[e^{ik\eta}(-i k\eta)[{\mathcal W}^{+}]_{\alpha\beta\gamma\delta}A - B e^{-ik\eta}(i k\eta)[{\mathcal W}^{-}]_{\alpha\beta\gamma\delta}]\label{meW2}\\
		& [{\mathcal W}^{+}]_{\alpha\beta\gamma\delta}^{\dagger} = [{\mathcal W}^{-}]_{\alpha\beta\gamma\delta}
	\end{align}
	which when evaluated gives exactly the result in Section \ref{TTTh}. Due to the form of (\ref{me}), the following contractions of $W$ appear in the time integral
	\begin{align}
		\mathcal{M}^{\pm\pm\pm}_{{\mathcal W}^3} = {\mathcal W}_{\alpha\beta\gamma\delta}^{\pm}(k_1){\mathcal W}_{\gamma\delta\eta\zeta}^{\pm}(k_2){\mathcal W}_{\eta\zeta\alpha\beta}^{\pm} (k_3)
	\end{align}
	where 
	\begin{align}
		{\mathcal W}^{\pm}_{\eta\zeta\alpha\beta}(k^{\mu}) \equiv {\mathcal W}_{\eta\zeta\alpha\beta}(\pm k, \vec{k})
	\end{align}
	However, only four of the sign combinations are unique
	\begin{align}
		&\mathcal{M}^{-++}_{{\mathcal W}^3} = \mathcal{M}^{+--}_{{\mathcal W}^3} \equiv \mathcal{M}^{2}_{{\mathcal W}^3}\\
		&\mathcal{M}^{+-+}_{{\mathcal W}^3} = \mathcal{M}^{-+-}_{{\mathcal W}^3}\equiv \mathcal{M}^{3}_{{\mathcal W}^3}\\
		&\mathcal{M}^{++-}_{{\mathcal W}^3} = \mathcal{M}^{--+}_{{\mathcal W}^3}\equiv \mathcal{M}^{4}_{{\mathcal W}^3}\\
		&\mathcal{M}^{---}_{{\mathcal W}^3} = \mathcal{M}^{+++}_{{\mathcal W}^3}\equiv \mathcal{M}^{1}_{{\mathcal W}^3}
	\end{align}
	The direct computation of the time integral (\ref{Tittt}) along with the shadow tranform gives the full homogenous part in momentum space,
	\begin{align}
		\langle TTT\rangle_{h, \alpha} = a\langle TTT\rangle_{h, 1}+b(\langle TTT\rangle_{h, 2}+\langle TTT\rangle_{h, 3}+\langle TTT\rangle_{h, 4})]\label{hgtttgr}
	\end{align}
	where $a, b$ are precisely what we 
	given by
	\begin{align}
		a&=\frac{1}{3\mathcal{N}^3(C, D)}[(2C^2+3 CD+3D^2)(\bar{C}^2+\bar{D}^2+\bar{C}\bar{D})+(C-D)\bar{D}(|C|^2+|D|^2+C\bar{D})]\nonumber\\
	    b&=\frac{1}{3\mathcal{N}^3(C, D)}[(C^2+6CD+ D^2)\bar{C}\bar{D}+C\bar{D}^2(-C+D)+(C-D)D\bar{C}^2]
	\end{align}
	We also have 
	\begin{align}
		\langle TTT\rangle_{h, i} = \langle TTT\rangle_{h, R_i}
	\end{align}
	where $\langle TTT\rangle_{h, R_i}$ appear in (\ref{homttt}), \eqref{homtttsh}. 
	\subsubsection*{Two-derivative interaction the Einstein-Hilbert contribution}
	Consider now the interaction 
	\begin{align}
		S^{(3)}_{\gamma, EG} = \int d^4x\sqrt{-g}~ R
	\end{align}
	The time integral due to the above interaction is then given by
	\begin{align}
		\mathcal{M}_{EG} \Im\bigg[&\int_{-\infty}^0\frac{d\eta}{\eta^2} \bigg(-2f_{k_3}(C, D)\gamma_{k_3}(\eta)[\bar{f}_{k_1}(C, D)\bar{f}_{k_2}(C, D) \bar{\gamma}_{k_1}(\eta)\bar{\gamma}_{k_2}(\eta)-2f_{k_1}(C, D)f_{k_2}(C, D)\gamma_{k_1}(\eta)\gamma_{k_2}(\eta)]\notag\\&+2\bar{f}_{k_1}(C, D)\bar{f}_{k_2}(C, D)\bar{\gamma}_{k_1}(\eta)\bar{\gamma}_{k_2}(\eta)[-\bar{f}_{k_3}(C, D) \bar{\gamma}_{k_3}(\eta)+f_{k_3}(C, D) \gamma_{k_3}(\eta)]\bigg)\bigg]
	\end{align}
	where
	\begin{align}
		\mathcal{M}_{EG} = (z_1.z_2 z_3.k_1+z_2.z_3 z_1.k_2+z_3.z_1 z_2.k_3)^2
	\end{align}
	Therefore, in momentum space, the full non-homogeneous part is then given by
	\begin{align}
		\langle TTT\rangle_{nh, \alpha} = a\langle TTT\rangle_{nh, 1}+b(\langle TTT\rangle_{nh, 2}+\langle TTT\rangle_{nh, 3}+\langle TTT\rangle_{nh, 4})]\label{Tnhttg1a}
	\end{align}
	where $a, b$ is precisely given by
	\begin{align}
		&a = \frac{1}{3\mathcal{N}^3(C, D)}[(2C^2+3 CD+3D^2)(C\bar{C}^3-D\bar{D}^3)+(C^3-D^3)(3\bar{C}^2\bar{D}+\bar{C}\bar{D}^2)] \notag\\
		 b&=\frac{1}{3\mathcal{N}^3(C, D)}[(C^2+6CD+ D^2)\bar{C}\bar{D}+C\bar{D}^2(-C+D)+(C-D)D\bar{C}^2]
	\end{align}
	\begin{align}
		\langle TTT\rangle_{nh, i} = \langle TTT\rangle_{nh, R_i}\quad i = 1, 2, 3, 4
	\end{align}
	where $\langle TTT\rangle_{nh, R_i}$'s appear in (\ref{othhno}).

	\newpage
	\bibliographystyle{JHEP}
	\bibliography{refs}

\end{document}